\documentclass[10pt, conference, compsocconf]{IEEEtran}

\newcommand{\ignore}[1]{}
\usepackage{fancyhdr}
\usepackage{epsfig}
\usepackage{graphicx} 
\usepackage{epstopdf}
\usepackage{setspace} 
\usepackage{subfig}
\usepackage[font=bf]{caption}
\usepackage{xcolor}
\usepackage{comment}
\usepackage{tikz}
\usepackage{url}
\usepackage{array}

\usepackage[pass]{geometry}
\usepackage[sort,nocompress]{cite}

\usepackage[normalem]{ulem}

\usepackage{fixltx2e}
\usepackage[ruled,vlined,linesnumbered]{algorithm2e}

\usepackage{flushend}

\newcommand{\beas}{\begin{eqnarray*}}
\newcommand{\eeas}{\end{eqnarray*}}
\newcommand{\bea}{\begin{eqnarray}}
\newcommand{\eea}{\end{eqnarray}}
\newcommand{\bes}{\begin{equation*}}
\newcommand{\ees}{\end{equation*}}
\newcommand{\be}{\begin{equation}}
\newcommand{\ee}{\end{equation}}

\newcommand\blfootnote[1]{%
  \begingroup
  \renewcommand\thefootnote{}\footnote{#1}%
  \addtocounter{footnote}{-1}%
  \endgroup
}

\newcommand*{\xdash}[1][3em]{\rule[0.5ex]{#1}{0.55pt}}

\newcommand*\redcircled[1]{\tikz[baseline=(char.base)]{
            \node[shape=circle,fill=red,font=\bfseries,inner sep=1pt] (char) {\textcolor{white}{#1}};}}

\iftrue
\newcommand{\nskim}[1]{{\color{red}[\textbf{\sc nskim}: \textit{#1}]}}
\else
\newcommand{\nskim}[1]{}
\fi

\newcommand{\newedit}[1]{{#1}}

\makeatletter
\def\BState{\State\hskip-\ALG@thistlm}
\makeatother

\makeatletter
\usepackage{xspace}
\def\@onedot{\ifx\@let@token.\else.\null\fi\xspace}
\DeclareRobustCommand\onedot{\futurelet\@let@token\@onedot}

\def\eg{\textit{e.g}\onedot} 
\def\ie{\textit{i.e}\onedot} 
\def\cf{\textit{cf}\onedot}

\def\arch{\texttt{CIAO}\xspace}
\makeatother

\begin{document}

\title{CIAO: Cache Interference-Aware Throughput-Oriented Architecture and Scheduling for GPUs} 

\author{\IEEEauthorblockN{Jie Zhang\IEEEauthorrefmark{1},
Shuwen Gao\IEEEauthorrefmark{2},
Nam Sung Kim\IEEEauthorrefmark{3}, 
and
Myoungsoo Jung\IEEEauthorrefmark{1}}
\IEEEauthorblockA{\IEEEauthorrefmark{1}School of Integrated Technology, Yonsei University, Korea}
\IEEEauthorblockA{\IEEEauthorrefmark{2}Intel, USA}
\IEEEauthorblockA{\IEEEauthorrefmark{3}University of Illinois Urbana-Champaign, USA}}

\maketitle


\begin{abstract}
A modern GPU aims to simultaneously execute more warps for higher  Thread-Level Parallelism (TLP) and performance. 
When generating many memory requests, however, warps contend for limited cache space and thrash cache, which in turn severely degrades performance. 
To reduce such cache thrashing, we may adopt cache locality-aware warp scheduling which gives higher execution priority to warps with higher potential of data locality.
However, we observe that warps with high potential of data locality 
often incurs far more cache thrashing or interference than warps with low potential of data locality. 
Consequently, cache locality-aware warp scheduling may undesirably increase cache interference and/or unnecessarily decrease TLP.

In this paper, 
we propose \textbf{C}ache \textbf{I}nterference-\textbf{A}ware throughput-\textbf{O}riented (\arch) on-chip memory architecture and warp scheduling
which exploit unused shared memory space and take insight opposite to cache locality-aware warp scheduling.
Specifically, \texttt{CIAO} on-chip memory architecture  can
adaptively redirect memory requests of severely interfering warps to unused shared memory space to isolate memory requests of these interfering warps from those of interfered warps.
If these interfering warps still incur severe cache interference, 
\texttt{CIAO} warp scheduling then 
begins to selectively 
throttle execution of these interfering warps. 
Our experiment shows that 
\texttt{CIAO} 
can offer 54\% higher performance than prior cache locality-aware scheduling at a small chip cost. 
\blfootnote{\xdash[18em] 

This paper is accepted by and will be published at 2018 International
Parallel and Distributed Processing Symposium. This document is presented to
ensure timely dissemination of scholarly and technical work.
}
\end{abstract}

\begin{IEEEkeywords}
GPGPU; thread-level parallelism; warp scheduling; interference;
\end{IEEEkeywords}

\section{Introduction}
\label{sec:introduction}
\noindent
The hardware-based warp scheduler of modern GPUs aims to schedule as many warps as possible to its Stream Multiprocessors (SMs) to increase TLP and thus performance~\cite{nvidia2009nvidia}.
Such warp scheduling, however, is not efficient for memory-intensive applications in which active warps collectively generate too many memory requests and thus contend for limited cache space~\cite{hong2009analytical, lee2012tap}.
Prior work reports that such cache contention (or interference) frequently incurs cache trashing and therefore severely degrades performances~\cite{jia2014mrpb, khairy2015efficient, tian2015adaptive}.
For example, our own experiment shows that a GPU can improve the geometric-mean performance of popular benchmark suites such as \texttt{PolyBench}~\cite{grauer2012auto}, \texttt{Mars}~\cite{he2008mars} and \texttt{Rodinia}~\cite{che2009rodinia} by 89\% when perfectly eliminating cache interference.

\begin{figure}[t]
\centering
\subfloat[]{\label{fig:colormap}\rotatebox{0}{\includegraphics[width=0.64\linewidth]{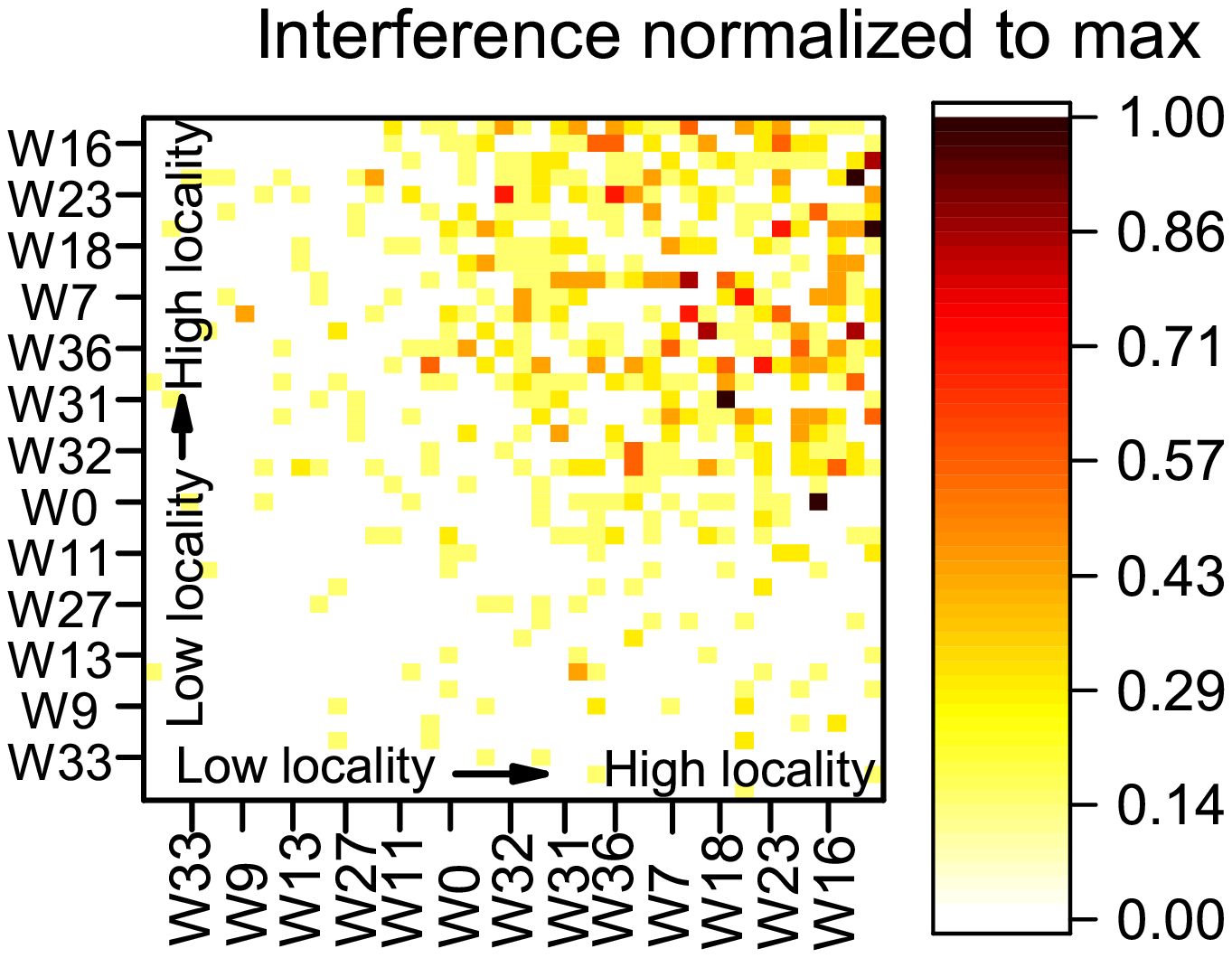}}}
\subfloat[]{\label{fig:moti_1}\rotatebox{0}{\includegraphics[width=0.36\linewidth]{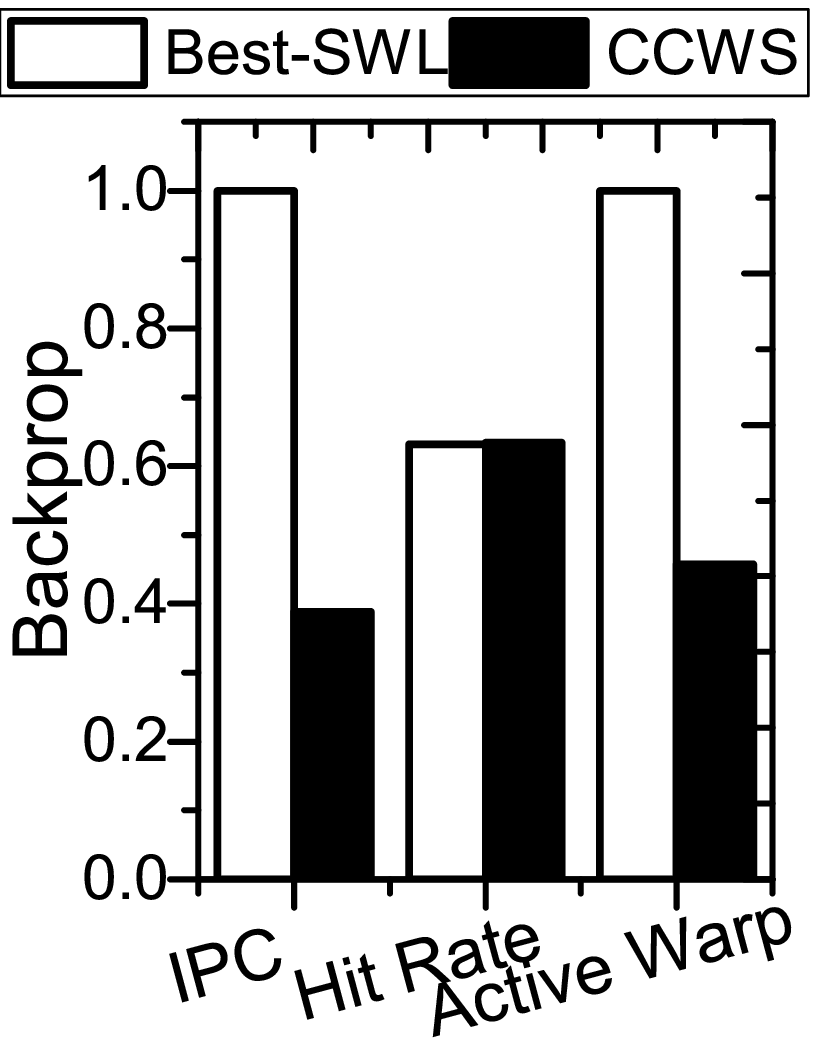}}}
\caption{\texttt{Backprop}~\cite{che2009rodinia}: (a) cache interference and (b) performance, cache hit rate, and number of active warps of \texttt{Best-SWL} and \texttt{CCWS}.}
\label{fig:motiv}
\end{figure}

To reduce cache thrashing, various cache-aware warp scheduling techniques have been proposed (\eg, \cite{jog2013owl, narasiman2011improving, rogers2012cache}).
These warp scheduling techniques aim to improve L1D cache hit rates and thus performance by identifying warps with high potential of data locality and then giving these warps a higher execution priority than other warps (\ie, judiciously reducing TLP by throttling execution of warps with low potential of data locality). 
However, we observe that such warp scheduling techniques are often inefficient especially for memory-intensive applications with \textit{irregular} cache access patterns 
due to two key reasons.
First, cache accesses of active warps with high potential of data locality often severely interfere with one another.
Consequently, scheduling warps simply based on their potential of data locality 
frequently increases cache interference with limited or even negative effect on improving cache hit rates. 
Second, it is undesirable to significantly reduce TLP in exchange for improved cache hit rates.
That is, although throttling execution of such warps may improve cache hit rates, overall performance can be either marginally improved or even degraded.

To provide better insights on our aforementioned observations, we run \texttt{Backprop}~\cite{che2009rodinia} using a popular GPU model~\cite{bakhoda2009analyzing}.
Then we analyze (1) which cache accesses of previously executed warps interfere with those of a currently executed warp and (2) how many cache misses of the currently executed warp are incurred by those of these previously executed warps in Figure~\ref{fig:colormap}.
This plot shows that a few warps with have high potential of data locality (\ie, \texttt{W16}, \texttt{W23}, and \texttt{W18}) also severely interfere with one another, incurring many unnecessary cache misses. 
We also consider two popular warp scheduling techniques: \texttt{Best-SWL} (Best-Static-Wavefront-Limiting scheduler)~\cite{rogers2012cache} and \texttt{CCWS} (Cache-Conscious Wavefront Scheduling)~\cite{rogers2012cache}, and compare performance, cache hit rate, and number of active warps of these two scheduling techniques in Figure~\ref{fig:moti_1}.
\texttt{Best-SWL} and \texttt{CCWS} aim to throttle execution of warps based on the best limitation value and potential of data locality determined by profiling and runtime techniques, respectively.
This plot shows that both \texttt{Best-SWL} and \texttt{CCWS} accomplish similar cache hit rates, but \texttt{Best-SWL} performs much better than \texttt{CCWS} as it reduces TLP less.

In this paper, tackling the aforementioned limitation of data locality-aware scheduling, we propose \textbf{C}ache \textbf{I}nterference-\textbf{A}ware throughput-\textbf{O}riented 
(\arch) on-chip memory architecture and warp scheduling which exploit unused shared memory space and take insight opposite to cache locality-aware scheduling. 
Specifically, we make the following contributions. 

First, we demonstrate that cache locality-aware scheduling, which gives higher execution priority to warps with \emph{higher} potential of data locality than warps with \emph{lower} potential of data locality, 
often undesirably increases cache interference and/or unnecessarily decreases TLP. 
Second, we propose \arch on-chip memory architecture that can adaptively redirect as many memory requests of interfering warps as possible to unused shared memory space, cost-effectively isolating memory requests of interfering warps from those of interfered warps.
This \arch on-chip memory architecture alone can notably reduce cache interference without diminishing TLP, providing {\bf 32}\% higher performance than \texttt{CCWS}.
Nonetheless, we may not be able to redirect every memory request of these interfering warps due to the limited unused shared memory space, and these warps may still incur severe cache interference.
To efficiently handle such a case, we then propose \arch warp scheduling which begins to selectively throttle execution of these interfering warps (\ie, giving lower priority to warps with \emph{high} potential of data locality), 
whereas \texttt{CCWS} throttles execution of warps with \emph{low} potential of data locality.
Lastly, the synergistic integration of \arch on-chip memory architecture and warp scheduling offers {\bf 54\%} higher performance than 
\texttt{CCWS}, because it significantly reduces L1D cache interference while keeping higher TLP. 

\section{Background}
\label{sec:background}
\subsection{GPU SM Architecture}
\label{sec:sm_arch}
\begin{figure}
\centering
\includegraphics[width=1\linewidth]{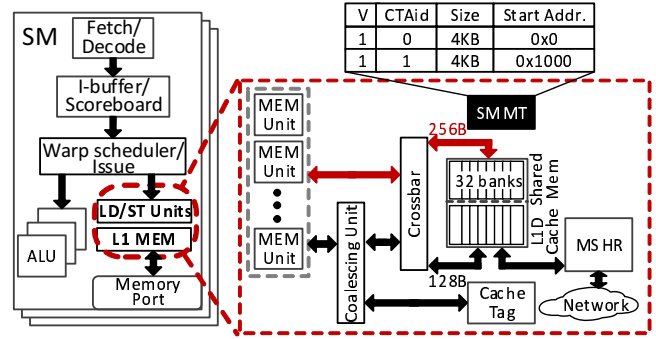}
\caption{GPU SM architecture.}
\label{fig:gpu1}
\end{figure}

\noindent  
\newedit{Figure~\ref{fig:gpu1} illustrates a representative SM architecture where shared memory may share a single on-chip memory structure with L1D cache~\cite{nvidia2009nvidia, nvidia2012nvidia}. 
The single on-chip memory structure consists of 32 banks with 512 rows, where 128 or 384 contiguous rows can be allocated to shared memory (\ie, 16KB or 48KB) based on user configuration and the remaining are allocated as L1D cache~\cite{gebhart2012unifying}.
While all 32 L1D cache banks operate in tandem for a single contiguous 32$\times$4-byte (128-byte) L1D cache request, all 32 shared memory banks can be accessed independently and serve upto 32 shared memory requests in parallel.
L1D cache buffers data from underlying memory and keeps a separate tag array to identify data hit. In such architecture, a L1D cache access is serialized. That is, tag array is accessed before the banks are accessed~\cite{edmondson1995internal}. 
In contrast, as shared memory stores intermediate results generated by ALU for each Cooperative Thread Array (CTA) which is explicitly manipulated by programmers, it neither needs tags nor accesses data in underlying memory. 
Hence, there is no datapath between shared memory and L2 cache, and no cache write/eviction policies are applied in shared memory~\cite{nvidia2009nvidia, nvidia2012nvidia}. In addition, to manage the shared memory space, each SM keeps an independent Shared Memory Management Table (SMMT)~\cite{yang2012shared}
where each CTA reserves one entry to store the size and base address of allocated shared memory. 
}

\begin{figure}[b]
\centering
\subfloat[]{\label{fig:l1_contention}\rotatebox{0}{\includegraphics[width=0.4\linewidth]{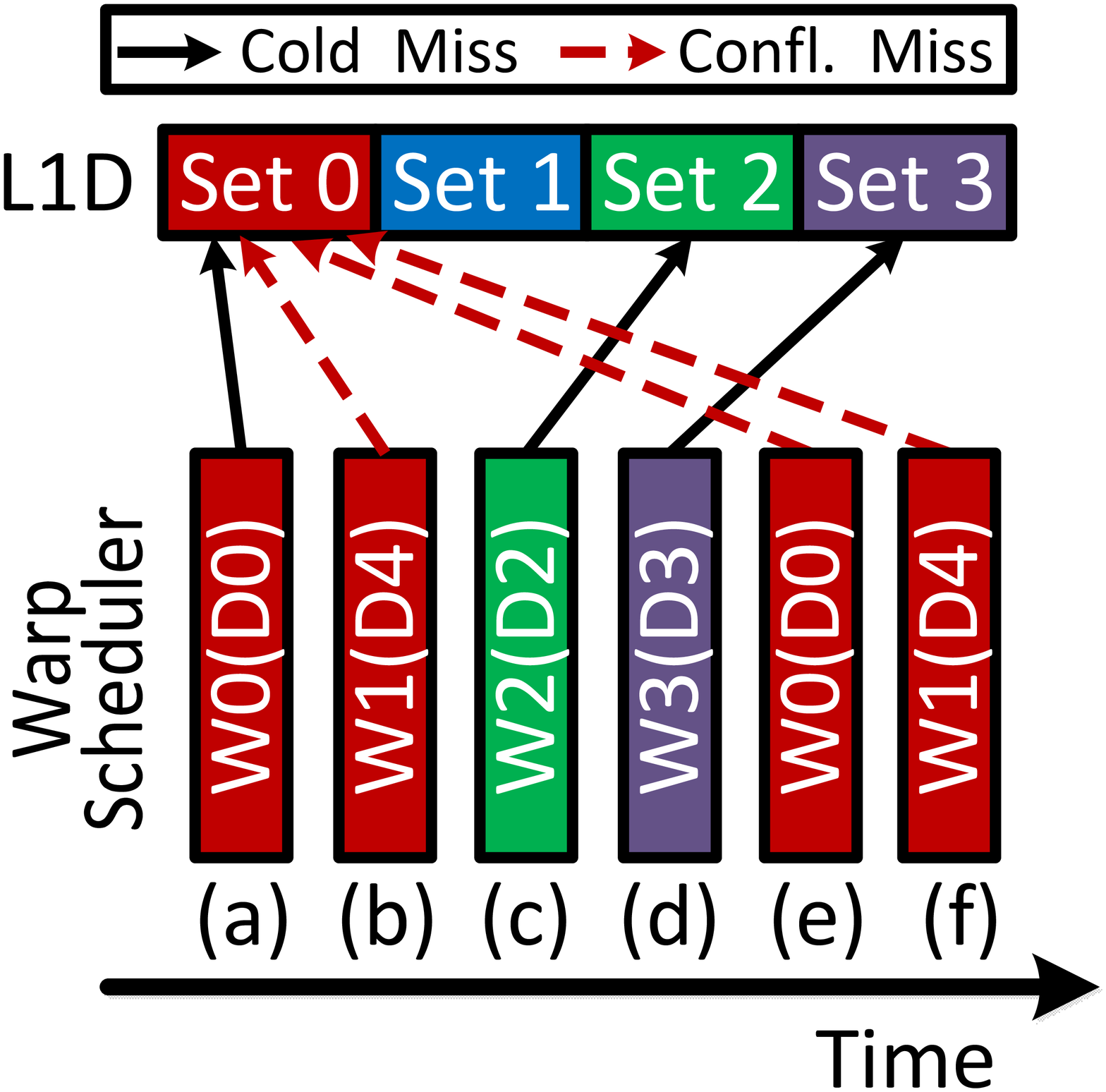}}}
\hspace{4pt}
\subfloat[]{\label{fig:VTA_mech}\rotatebox{0}{\includegraphics[width=0.57\linewidth]{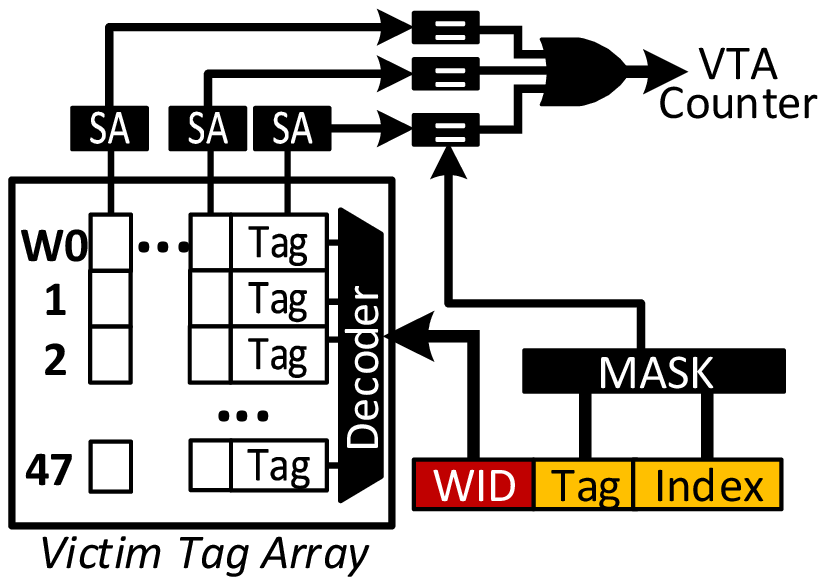}}}
\caption{(a) An example of locality and interference and (b) VTA structure.}
\end{figure}

\subsection{Cache Interference}
\label{sec:interfere}
\noindent 
As many warps share small L1D cache, they often contend for the same cache line.
Hence, cached data of an active warp are frequently evicted by cache accesses of other active warps. 
This phenomenon is referred to as \textit{cache interference} which often changes supposedly a regular memory access pattern into an irregular one. 
Figure~\ref{fig:l1_contention} depicts an example of how the cache interference worsens data locality in L1D cache, 
where warps \texttt{W0} and \texttt{W1} send memory requests to get data \texttt{D0} and \texttt{D4}, respectively.
However, since \texttt{D0} and \texttt{D4} are mapped to the same cache set \texttt{S0}, 
repeated memory requests from \texttt{W0} and \texttt{W1} to get \texttt{D0} and \texttt{D4} keep evicting \texttt{D4} and \texttt{D0} at cycles \texttt{(a)}, \texttt{(b)}, \texttt{(e)}, and \texttt{(f)}.
Unless the memory requests from \texttt{W1} and \texttt{W0} evicted \texttt{D0} and \texttt{D4}, respectively, 
they should have been L1D cache hits.
Such a cache hit opportunity is also called \textit{potential of data locality}, which can be quantified by the frequency of re-referencing the same data unless cache interference occurs.

\subsection{Potential of Data Locality Detection}
\label{sec:vta}
\noindent
To detect the potential of data locality described in Section~\ref{sec:interfere}, we may leverage a Victim Tag Array (VTA)~\cite{rogers2012cache} where
we store a Warp ID (WID) in each cache tag, as shown in Figure~\ref{fig:VTA_mech}.
A WID in a cache tag is to track which warp brought current data in a cache line. 
When a memory request of a warp evicts data in a cache line, we first take 
(1) the address in the cache tag associated with the evicted data and 
(2) the WID of the warp evicting the data. 
Then we store (1) and (2) in a VTA entry which is indexed by the WID stored in the cache tag (\ie, the WID of the warp which brought the evicted data in the cache line).
When memory requests of an active warp repeatedly incur VTA hits, 
they exhibit potential of data locality.

\section{Architecture and Scheduling}
\label{sec:overview}
\noindent
In this section, we overview \texttt{CIAO} (1) cache interference detection mechanism; (2) on-chip memory architecture and (3) warp scheduling,
which can synergistically reduce cache thrashing without notably hurting TLP. 
We will describe their implementation details in Section~\ref{sec:implementation}.

\subsection{Cache Interference Detection}
\label{sec:interference_detection}
\noindent
As introduced in Section~\ref{sec:interfere}, some warps incur more severe cache interference than other warps (\ie, non-uniform cache interference). 
However, it is non-trivial to capture such non-uniform interference occurring during the execution of applications 
at compile time \cite{chenadaptive}.
Thus, we need to determine severely interfering and interfered warps at runtime.

At run time, we may track severely interfered warps, leveraging a VTA structure (\cf Section~\ref{sec:vta}).
A na\"ive way to determine severely interfering warps for each warp, however, demands a high storage cost, 
because each warp needs to keep track of cache misses incurred by all other $n-1$ warps.
This in turn requires a storage structure with $n(n-1)$ entries where $n$ is the number of active warps per SM (\ie, 48 warps). 
Searching for a cost-effective way to determine severely interfering warps, we exploit our following observation on an important characteristic of cache interference. 

\begin{figure}
\centering
\subfloat[]{\label{fig:unbalance_arrow}\rotatebox{0}{\includegraphics[width=0.34\linewidth]{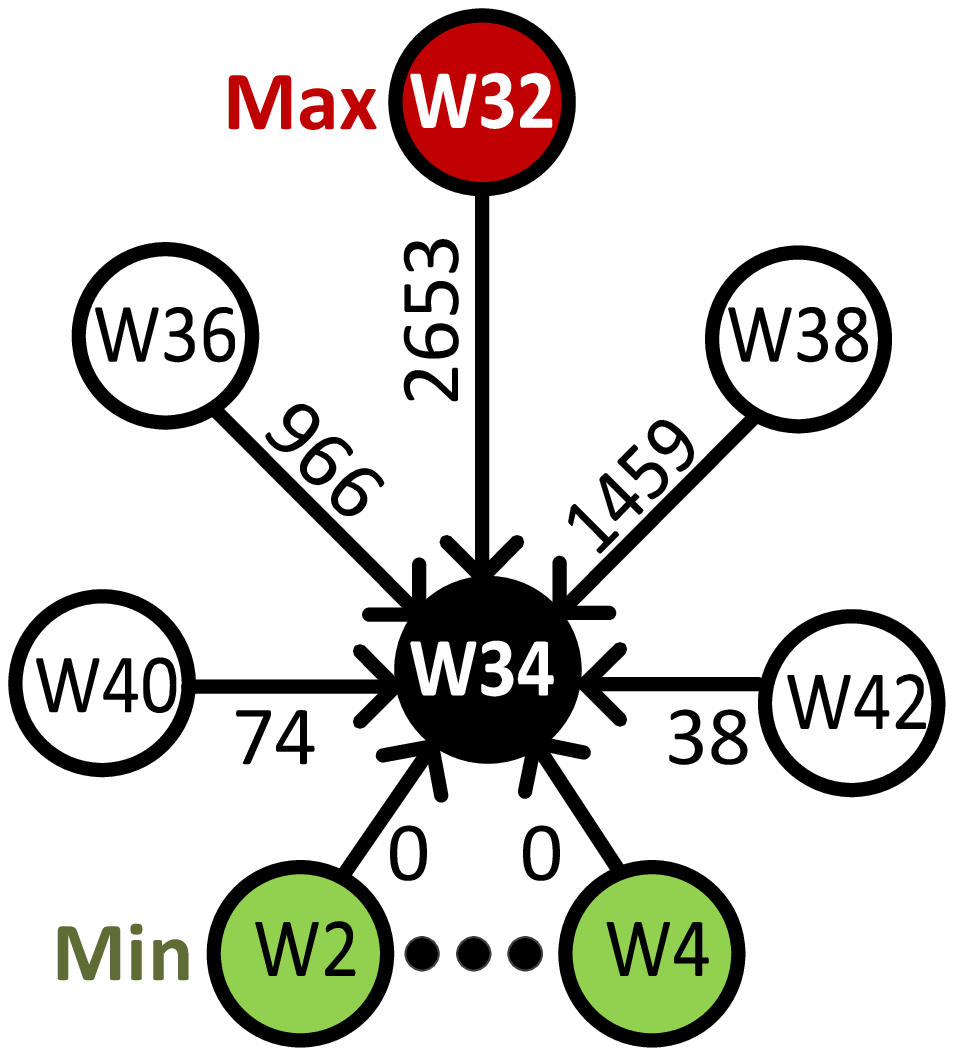}}}
\subfloat[]{\label{fig:kmeans_un1}\rotatebox{0}{\includegraphics[width=0.61\linewidth]{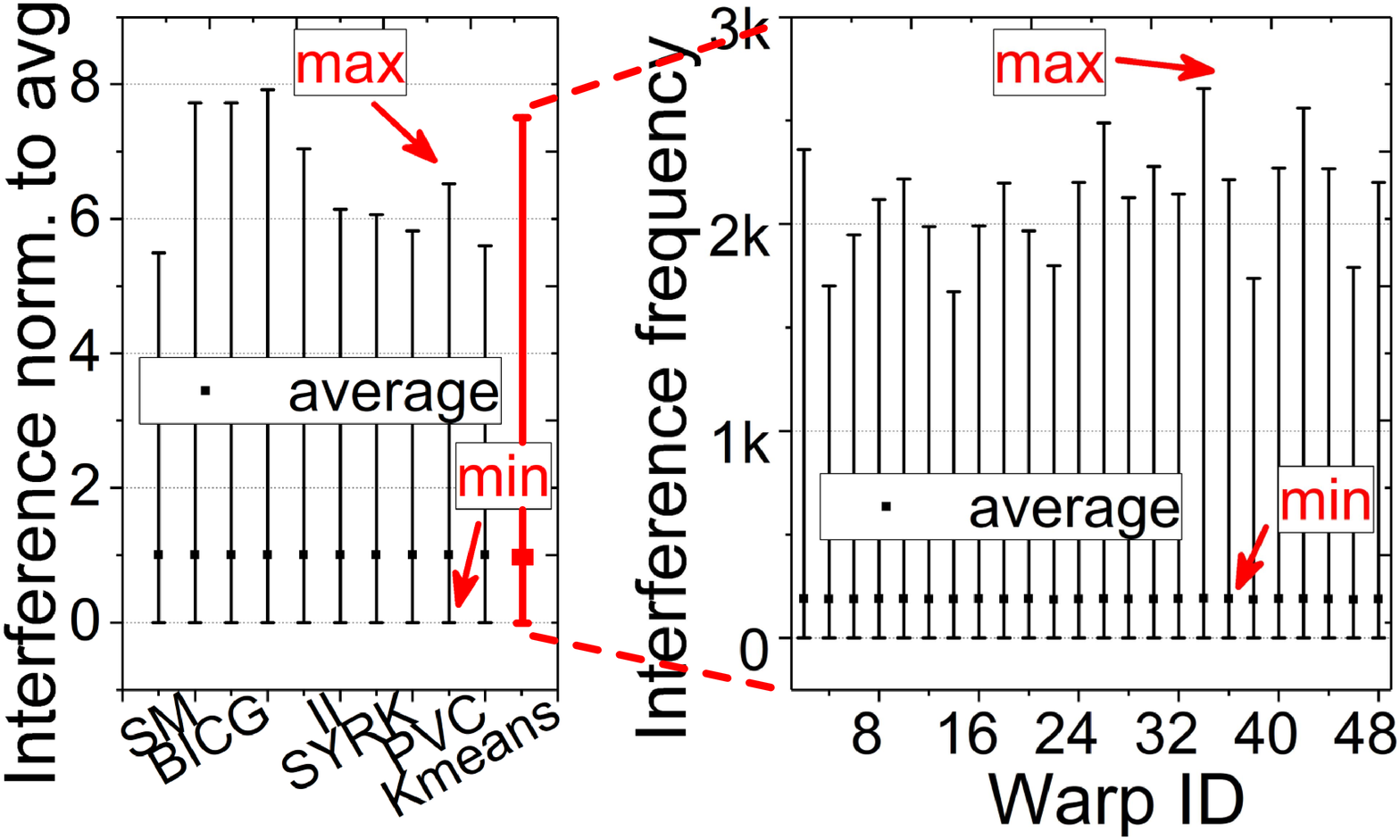}}}

\subfloat[]{\label{fig:sat_counter}\rotatebox{0}{\includegraphics[width=0.92\linewidth]{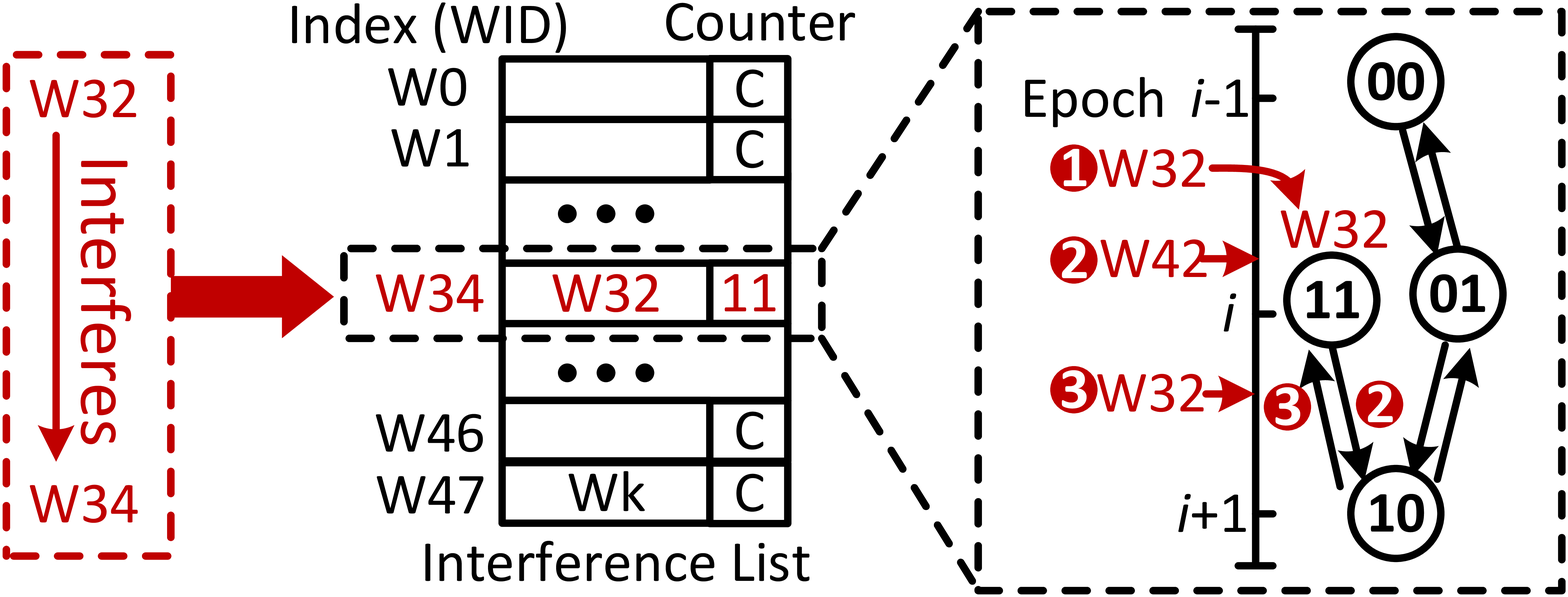}}}
\caption{(a) Warps interfering with warp W34 and their interference frequency.
(b)  Min and max interference frequencies experienced by each warp and each evaluated workload.  (c) Interference detection example.}
\end{figure}

\begin{figure*}
\centering
\includegraphics[width=1\linewidth]{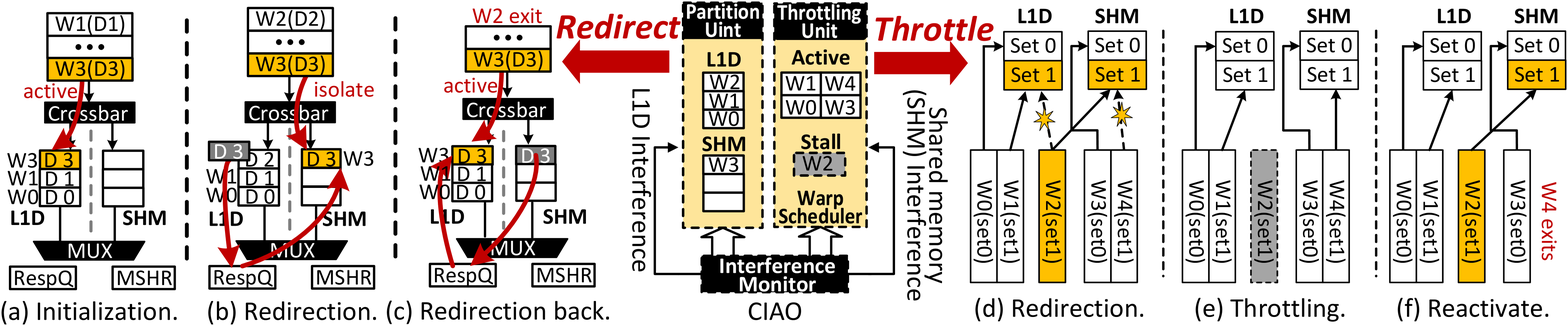}
\caption{\arch execution flow.}
\label{fig:thrott_exam}
\end{figure*}

\newedit{ Figure~\ref{fig:unbalance_arrow} shows that \texttt{W32} interferes with \texttt{W34}, more than two thousand times, whereas some warps (\eg, \texttt{W2}) do not interfere with \texttt{W34} at all in \texttt{KMEANS}~\cite{che2009rodinia}; 
we observe a similar trend on cache interference in all other benchmarks that we tested (cf. Figure~\ref{fig:kmeans_un1}). }
Observing such an interference characteristic, we propose to track only the most recently and frequently interfering warp for each warp.
This significantly reduces the storage cost required to track every interfering warp for each warp.
Specifically, \arch keeps a small memory structure denoted by \textit{interference list} 
where each entry is indexed by the WID of a currently executed warp.

To track the most recently and frequently interfering warp for a currently executed warp, 
we may augment each list entry with a 2-bit saturation counter.
Figure~\ref{fig:sat_counter} illustrates how \arch utilizes the counter to track an interfering warp. 
Suppose that a previously executed warp (\texttt{W32}) interfered with a currently executed warp (\texttt{W34}), 
That is, \texttt{W32} is an interfering WID and \texttt{W34} is an interfered WID.
Subsequently, the interfering WID is stored in the list entry indexed by the interfered WID,
and the counter in the list entry is set to \texttt{00}; 
the interfering WID is provided by a VTA entry field that tracks which warp incurred the last eviction (\cf Section~\ref{sec:vta}).

Whenever \texttt{W32} interferes with \texttt{W34} (not shown in the figure), the counter is incremented by 1.
Suppose that the counter has already reached \texttt{11} (\redcircled{\small{1}}) at a given cycle. 
When another warp (\texttt{W42}) interferes with warp \texttt{W34} in a subsequent cycle, the counter is decremented by 1 (\redcircled{\small{2}}). 
Then, if warp \texttt{W32} interferes with \texttt{W34} again, the counter is incremented by 1 (\redcircled{\small{3}}). 
The interfering WID in the list entry is replaced with the most recent interfering WID only when its saturation counter is decreased to \texttt{00}, 
so that the warp with most frequent cache interference can be kept in the interference list.

\subsection{CIAO On-Chip Memory Architecture}
\label{sec:warp_partitioning}
\noindent
An effective way to reduce cache interference is to isolate cache accesses of interfering warps from those of interfered warps 
after partitioning the cache space and allocating separate cache lines to the interfering warps. 
Prior work proposed various techniques to partition the cache space for CPUs (\eg, \cite{qureshi2006utility, srikantaiah2008adaptive}). 
However, the size of L1D cache is insufficient to apply such techniques for GPUs, 
as the number of GPU threads sharing L1D cache lines is very large, compared with that of CPU threads. 
For example, only two or three cache lines can be allocated to each warp, if we apply a CPU-based cache partitioning technique to the L1D cache of GTX480. 
Such a small number of cache lines per warp can even worsen cache thrashing.

\newedit{Meanwhile, we observe that programmers prefer L1D cache rather than shared memory for programming simplicity and the limited number of running GPU threads constrains the usage of shared memory, leading to 
a large fraction of shared memory unused 
(\cf $F_{smem}$ of Table~\ref{tab:workload_charac} in Section~\ref{sec:method}).
This agrees to prior work's analysis~\cite{hayes2014unified, virtualthread}. }
Exploiting such unused shared memory space, 
we propose to redirect memory requests of severely interfering warps to the unused shared memory space.

As there is no cache interference at the beginning of kernel execution, 
memory requests of all the warps are directed to L1D cache, as depicted in Figure~\ref{fig:thrott_exam}a. 
However, as the kernel execution progresses, cache accesses begin to compete one another to acquire specific cache lines in L1D cache. 
As the intensity of cache interference exceeds a threshold, \arch determines severely interfering warps (\cf Section~\ref{sec:interference_detection}).
Subsequently, \arch redirects memory requests of these interfering warps to unused shared memory space, 
isolating the interfering warps from the interfered warps in terms of cache accesses, as depicted in Figure~\ref{fig:thrott_exam}b.
This in turn can significantly reduce cache contentions without throttling warps (\ie, hurting TLP). 
\newedit{After the redirection, the memory requests are forwarded from L1D cache to shared memory but the data may already present in the L1D cache (\cf W3/D3 in Figure~\ref{fig:thrott_exam}b). To guarantee cache coherence between L1D cache and shared memory, single data copy needs to be exclusively stored in either shared memory or L1D cache. 
Such challenge can be addressed by migrating the data copy from L1D cache to shared memory, which may take the steps as follows: 1) a data miss signal would be raised for shared memory, 2) the data copy in L1D cache would be evicted to response queue, and 3) a new entry of MSHR would be filled with the pointer referring to the location of single data copy in the response queue. Later on, to fill the data miss, shared memory fetches data from response queue based on the location information recorded in MSHR.
}
When \arch detects significant decrease in cache contentions due to a change in cache access patterns or completion of execution of some warps, 
it redirects the memory requests of these interfering warps from shared memory back to L1D cache (\cf Figure~\ref{fig:thrott_exam}c).

To exploit the unused shared memory space for the aforementioned purpose, however, there are two challenges. 
First, the shared memory has its own address space separated from the global memory, 
and there is no hardware support that translates a global memory address to a shared memory address. 
Second, the shared memory does not have a direct datapath to L2 cache and main memory~\cite{jamshidi2014d}. 
That is, it always receives and sends data only through the register file.
To overcome these limitations, we propose to adapt shared memory architecture as follows. 
First, we implement a address translation unit in front of shared memory to translate a given global memory address to a local shared memory address.
Second, we slightly adapt the datapath between L1D and L2 caches such that the shared memory can also access L2 cache when the unused shared memory space serves as cache.

\subsection{CIAO Warp Scheduling}
\label{sec:warp_throttling}
\noindent
Although \arch on-chip memory architecture 
can effectively isolate cache accesses of interfering warps from those of interfered warps, its efficacy depends on various run-time factors, such as the number of interfering warps and the amount of unused shared memory space. 
For example, 
the interfering warps end up thrashing the shared memory as well when the amount of unused shared memory space 
is insufficient to handle a large number of memory requests from the interfering warps in a short time period (cf. Figure~\ref{fig:thrott_exam}d).

To efficiently handle such a case, we propose to throttle interfering warps \textit{only} when it is not effective to redirect memory requests of interfering warps to the shared memory.
Specifically, sharing the same cache interference detector used for \arch on-chip memory architecture, 
\arch monitors the intensity of interference at the shared memory at runtime.
Once the intensity of interference at the shared memory exceeds a threshold, \arch stalls
the most severely interfering warp at the shared memory (\eg, \texttt{W2} in Figure~\ref{fig:thrott_exam}e). 
\arch repeats this step until the intensity of interference at the shared memory falls below the threshold. 
As some warps complete their execution and subsequently the intensity of interference at the shared memory falls below the threshold,
\arch starts to reactivate the stalled warp(s) in the reverse order to keep high TLP and maximize the utilization of shared memory (\cf Figure~\ref{fig:thrott_exam}f).

Note that \arch warp scheduling shares the same interference detector with \arch on-chip memory architecture, instead of 
keeping two separate interference detectors for L1D and shared memory, respectively. 
This is because isolated interfering warps do not compete L1D cache with warps that exclusively access L1D cache, and memory accesses of isolated interfering warps often interfere with one another. 
In other words, L1D cache and shared memory interferences do not affect each other. 
Hence, L1D cache and shared memory can share the same VTA array to detect interferences.

\section{Implementation}
\label{sec:implementation}
\noindent
In this section, we present required GPU microarchitecture adaptations to implement the interference detector, on-chip memory architecture, and warp scheduling. 

\subsection{Cache Interference Detection}
\label{sec:schedule}

\noindent \textbf{Estimation of cache interference.}
A level of cache interference experienced by a warp can be quantified by 
an Individual Re-reference Score (IRS) 
which can be expressed by:

\begin{equation}
\label{eq:irs}
IRS_i = \frac{F^i_{VTA-hits}}{N_{executed-inst}/N_{active-warp}}
\end{equation}

\begin{figure}[b]
\centering
\includegraphics[width=1\linewidth]{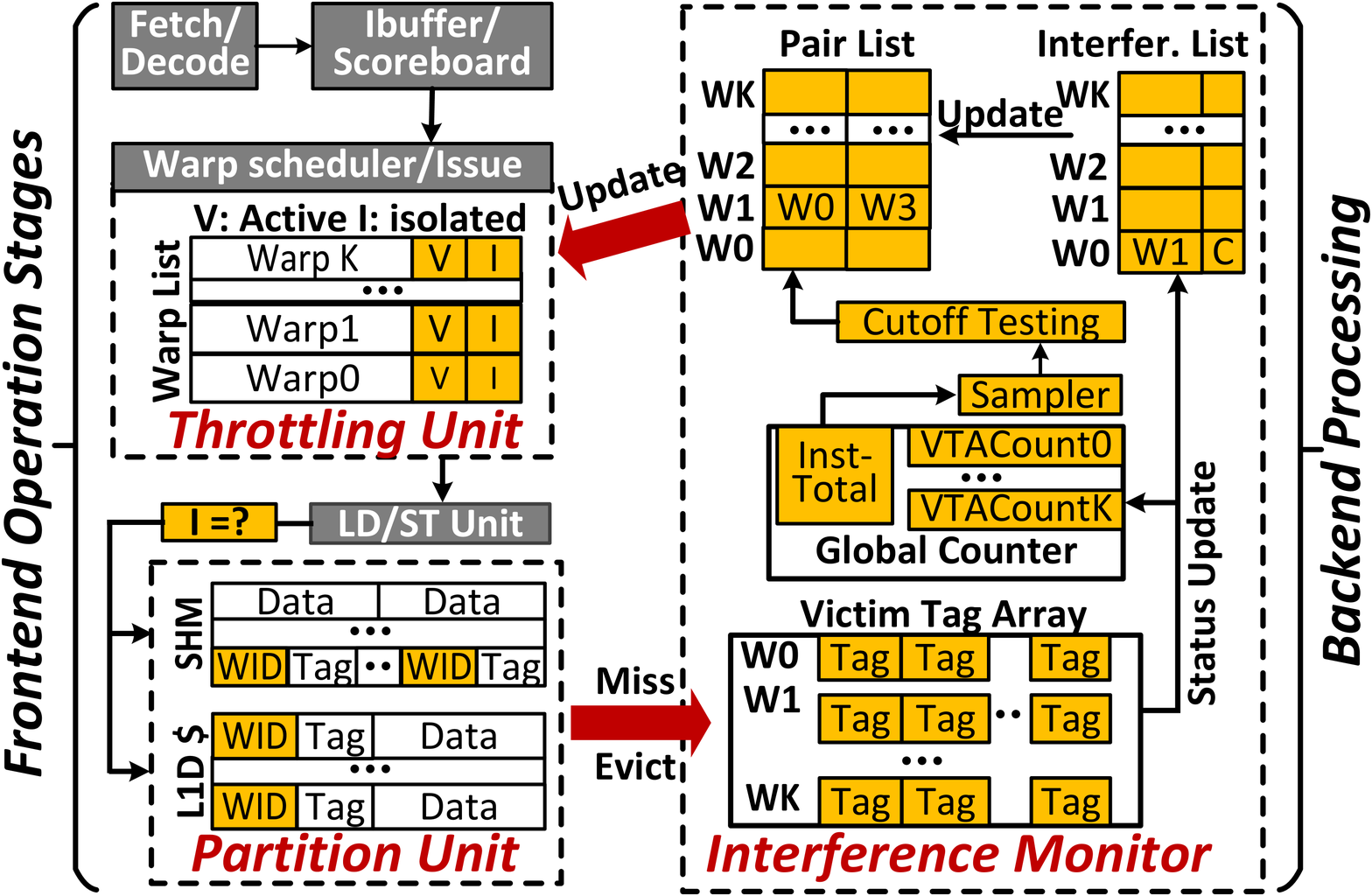}
\caption{Microarchitecture adaptation for \arch.}
\label{fig:vta}
\end{figure}

\noindent where $i$ is active warp number, $F^i_{VTA-hits}$ is the number of VTA hits for warp $i$, $N_{executed-inst}$ is the total number of executed instructions, and $N_{active-warp}$ is the number of active warps running on an SM, respectively. 
$IRS_i$ represents VTA hits per instruction (\ie, intensity of VTA hits) for warp $i$.
High $IRS_i$ indicates warp $i$ has experienced severe cache interference in a given epoch.
Based on $IRS_i$, \arch (1) decides whether it isolates warps interfering with warp $i$, (2) stalls these interfering warps, or (3) reactivates the stalled warps.

\begin{figure*}
\centering
\includegraphics[width=1\linewidth]{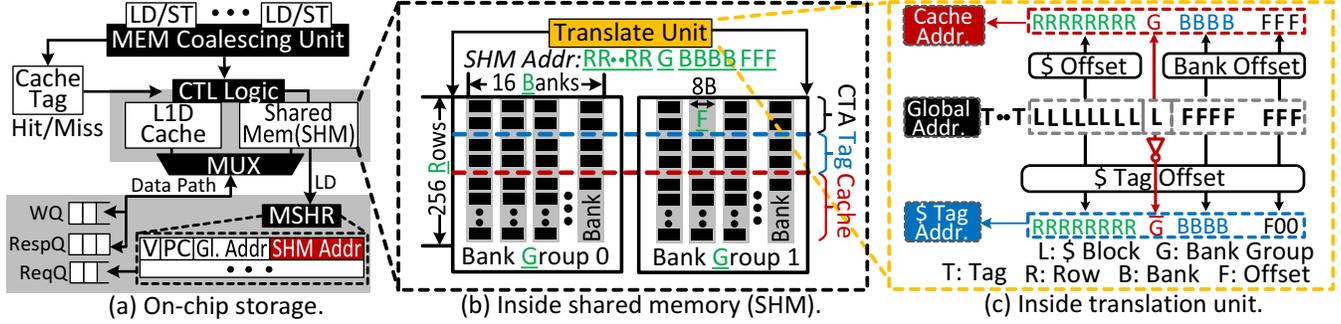}
\caption{GPU on-chip memory structure adaptation.}
\label{fig:shm}
\end{figure*}

\noindent \textbf{Decision thresholds.} 
For these aforementioned three decisions we introduce two threshold values: (1) \texttt{high-cutoff} and (2) \texttt{low-cutoff}.  
$IRS_i$ over \texttt{high-cutoff} indicates that warp $i$ has experienced severe cache interference.
Subsequently, \arch decides to isolate or stall the warp that most recently and severely interfered with warp $i$. 
$IRS_i$ below \texttt{low-cutoff} often indicates that warp $i$ has experienced light cache interference and/or completed its execution.
Then, \arch decides to reactivate previously stalled warps or 
redirect memory requests of these warps back to L1D cache.
As these two thresholds influence the efficacy of \arch, we sweep these two values, evaluate diverse memory-intensive applications, and determine that \texttt{high-cutoff} and \texttt{low-cutoff}, which minimize cache interference and maximize performance, are 0.01 and 0.005, respectively.
See Section~\ref{subsubsec:sensi} for our sensitivity analysis.

\noindent \textbf{Epochs.} 
As $IRS_i$ changes over time, \arch should track the latest $IRS_i$ and compare it against \texttt{high-cutoff} and \texttt{low-cutoff} to precisely determine whether a warp needs to be isolated, stalled, or reactivated. However, the update of $IRS_i$ calculation consumes more than 6 cycles, which can be on the critical path of performance.
To this end, \arch divides the execution time into \texttt{high-cutoff} and \texttt{low-cutoff} epochs, respectively. 
At the end of each \texttt{high-cutoff} (or \texttt{low-cutoff}) epoch, \arch updates $IRS_i$ and compares it against \texttt{high-cutoff} (or \texttt{low-cutoff}).
The \texttt{low-cutoff} epoch should be shorter than the \texttt{high-cutoff} epoch because of the following reasons. 
As preserving high TLP is a key to improve GPU performance, \arch attempts to minimize a negative effect of stalling warps by reactivating stalled warps as soon as these warps start not to notably interfere with other warps at runtime.  
To validate this strategy, we sweep \texttt{high-cutoff} and \texttt{low-cutoff} epoch values, evaluate diverse memory-intensive applications, and determine that the best \texttt{high-cutoff} and \texttt{low-cutoff} epoch values are every 5000 and 100 instructions, respectively.
See Section~\ref{subsubsec:sensi} for our in-depth sensitivity analysis.

\noindent \textbf{Microarchitecture support.}
Figure~\ref{fig:vta} depicts the necessary hardware, which is built upon the existing VTA organization~\cite{rogers2012cache}, to implement a cache interference detector.

To capture different levels of cache interference experienced by individual warps, we implement a VTA-hit counter per warp and a total instruction counter per SM (\texttt{VTACount0-k} and \texttt{Inst-total} in the figure) atop a VTA.
Each VTA-hit counter records the number of VTA hits for each warp, and the total instruction counter tracks the total number of instructions executed by a given SM (\ie, $N_{executed-inst}$ in Eq.(\ref{eq:irs})). 
To compare $IRS_i$ against \texttt{high-cutoff} and \texttt{low-cutoff}, we implement the cutoff testing unit which can be implemented by registers, a shifter, and simple comparison logic. 
Lastly, we implement the samplers to count the number of executed instructions and determine whether or not the end of a \texttt{high-cutoff} or \texttt{low-cutoff} epoch has been reached.

To manage the information related to tracking interfering warps for each warp, we implement the interference list.
Each entry is indexed by WID of a given warp and stores a 6-bit WID of an interfering warp and a 2-bit saturation counter (\texttt{C} in the figure). 
When a VTA hit occurs, the corresponding entry of interference list is updated, as described in Section~\ref{sec:interference_detection}.
\arch checks the interference list for warp $i$ whenever it needs to isolate or stall an interfering warp based on $IRS_i$.
To facilitate this, we also augment a 1-bit active flag (\texttt{V}) and 1-bit isolation flag (\texttt{I}) with each ready warp entry in the warp list (\ie, a component of warp scheduler). 
Using \texttt{V} and \texttt{I} bits, the warp scheduler can identify whether a given warp is in active (\texttt{V}=\texttt{1}, \texttt{I}=\texttt{0}), isolated (\texttt{V}=\texttt{1}, \texttt{I}=\texttt{1}), or stalled state (\texttt{V}=\texttt{0}).

We also implement a \textit{pair list}. 
Each entry is indexed by the WID of a warp at the front of the warp list and composed of two fields 
to record which interfered warp triggered to redirect memory requests of the warp or stall the warp in the past.
Suppose that warp $i$ is at the front of the warp list.
Based on WIDs from the first or second field of the entry indexed by warp $i$, 
\arch checks $IRS_k$ where $k$ is the WID of the interfered warp that previously triggered to either redirect memory requests of warp $i$ or stalling warp $i$. 
Then \arch decides whether it reactivates warp $i$ or redirects memory requests of warp $i$ back to L1D cache based on $IRS_k$. 
For example, as \texttt{W0} is severely interfered by \texttt{W1}, \arch decides to redirect memory requests of \texttt{W1} to unused shared memory space.
Then \texttt{W0} is recorded in the first field of the entry indexed by \texttt{W1} and \texttt{I} associated with \texttt{W1} is set, as depicted in Figure~\ref{fig:vta}. 
Subsequently, \texttt{W1} begins to send memory requests to the shared memory, but \arch observes that \texttt{W1} also severely interferes with \texttt{W3} that sends its memory requests to the shared memory.
As \arch decides to stall \texttt{W1}, \texttt{W3} is recorded in the second field of the entry indexed by \texttt{W1} and \texttt{V} associated with \texttt{W1} is cleared. 
When \arch needs to reactivate \texttt{W1} later, the second field of the pair list entry and \texttt{V} corresponding to \texttt{W1} are cleared to 
inform the warp scheduler of the event that the warp is active. 
When \arch needs to make \texttt{W1} send its memory request back to L1D cache, the corresponding field in the pair list entry and \texttt{I} are cleared.
See Section~\ref{sec:pat} for more details on the pair list.

\subsection{Shared Memory Architecture}
\label{sec:shared_mem_arch}
\noindent
Figure~\ref{fig:shm}a and b illustrate \arch on-chip memory architecture and its data placement layout, respectively.

\noindent \textbf{Determination of unused shared memory space.}
One challenge to utilize unused shared memory space is that shared memory is managed  by programmers and the used amount of shared memory space varies across implementations of a kernel. 
To make \arch on-chip memory architecture transparent to programmers, we leverage the existing SMMT structure to determine the unused shared memory space. 
When a CTA is launched, \arch checks the corresponding SMMT entry to determine the amount of unused shared memory space (\cf Section~\ref{sec:sm_arch}). 
Then, \arch inserts a new entry in the SMMT with the start address and size of unused shared memory to reserve the space for storing 128-byte data blocks and tags.

\noindent \textbf{Placement of tags and data.}
In contrast to L1D cache, shared memory does not have a separate memory array to accommodate tags~\cite{gebhart2012unifying}. 
In this work, instead of employing an additional tag array, we propose to place both 128-byte data blocks and their tags into the shared memory.
This is to minimize the modification of the current on-chip memory structure architected to be configured as both L1D cache and shared memory. 
As shown in Figure~\ref{fig:shm}b, we partition 32 shared memory banks into two bank groups and stripe a 128-byte data block across 16 banks within one bank group. 
Each 128-byte data block can be accessed in parallel since each shared memory bank allows 64-bit accesses~\cite{nvidia2012nvidia}.
Since a tag and a WID require only 31 bits (= 25 + 6 bits), two tags can be placed in a single bank which is different from banks storing the corresponding data blocks.
Then 32 tags can be grouped together to better utilize a row of one bank group (\ie, 16 banks). 
This design strategy, which puts a tag and the corresponding data block into two different bank groups, shuns bank conflicts and thus allows accesses of a tag and a data block in parallel.
Furthermore, we only use the unused shared memory space as direct-mapped cache so that a pair of a 128-byte data block and the corresponding tag can be accessed with a single shared memory access.

\noindent \textbf{Address translation unit.}
As shown in Figure~\ref{fig:shm}b, we introduce a hardware address translation unit in front of shared memory to determine where a target 128-byte data block and its tag exist in the shared memory. 
In practice, a global memory address can be decomposed by cache-related information such as a tag, block index and byte offset. 
However, as the usage of shared memory can be varying based on the needs of each CTA, we put an 8-bit mask into the translation unit to decide how many rows will be used for each CTA at runtime. 
Figure~\ref{fig:shm}c shows how our translation unit determines locations of a target data block and its tag; 
the data block address (of shared memory) consists of four fields, the byte offset (``\texttt{F}''), bank index (``\texttt{B}''), bank group (``\texttt{G}''), and row index (``\texttt{R}''), which are presented from LSB to MSB.  
Specifically, we have 8-byte rows per bank, 16 banks per group, two bank groups and 256 rows (at most), 
which in turn 3, 4, 1, and 8 bits for \texttt{F}, \texttt{B}, \texttt{G} and \texttt{R}, respectively. 
The remaining bits (16 bits in this example) are used as part of the tag. 
Note that our tags also contain 6-bit WID and 9-bit data block index as the number of cache lines required can be greater than the number of rows. 

In \arch, one row within a bank group can hold 32 tags since a physical row per bank contains two tags. 
That is, the actual position of a tag can be indicated by 5 bits (\ie, 1 \texttt{F} and  4 \texttt{B} bits), 
which are also used for the row index of the corresponding data block. 
To access a data block and the corresponding tag in parallel, \texttt{G} of the data block will be flipped and assigned to such tag's 5 bits as a significant bit. 
The remaining \texttt{R} bits are assigned to the row index of the target tag. 
Note that, as shown in the figure, the start of index for both a data block and a tag can be rearranged by considering the data block and tag offset registers, 
which are used to adapt the unused shared memory size allocated for cache.

\noindent \textbf{Datapath connection.}
When we leverage unused shared memory as cache, we need a datapath between shared memory and L2 cache. 
Since the shared memory is disconnected from the global memory in the conventional GPU, 
we need to adapt the on-chip memory structure, which is partitioned between L1D cache and shared memory, to share some resources of the L1D cache with the shared memory (\eg, datapath to L2 cache, MSHR, etc.). 
As illustrated in Figure~\ref{fig:shm}a, 
a multiplexer is implemented to connect the write queue (WQ) and response queue (RespQ) to either L1D cache or shared memory. 
The \arch cache control logic controls the multiplexer based on the isolation flag bit (\texttt{I}) and the result of checking cache tags associated with accessing L1D cache or shared memory serving as cache.
We also augment an extra field with each MSHR entry to store the shared memory address of a memory request from the aforementioned address translation unit. 
Once the shared memory issues a fill request after a miss, the request reserves one MSHR entry by filling in its global and translated shared memory addresses. 
If the response from L2 cache matches the global address recorded in the corresponding MSHR entry, the filling data can be directly stored in the shared memory based on the translated shared memory address.

\noindent \textbf{Performance optimization and coherence.}
When \arch redirects memory requests of an interfering warp from L1D cache to shared memory, the shared memory does not have any data. 
This can incur (1) performance degradation because of cold misses and (2) some coherence issues.
To address these two issues, 
when \arch needs to access the shared memory, the cache controller first checks the tag array of L1D cache.
If a target data resides in L1D cache (not in shared memory), the L1D cache will evict the data directly to the response queue, 
which is used to buffer the fetched data from L2 cache and invalidate the corresponding cache line in L1D cache. 
Note that checking the tag array and accessing L1D cache are serialized as described in Section~\ref{sec:sm_arch}. 
Meanwhile, the shared memory issues a fill request to MSHR, as the shared memory does not have the data yet. 
During this process, the target data will be directly fetch from the response queue to the shared memory (\cf Figure~\ref{fig:shm}a) 
In this way, we naturally migrate data from L1D cache to shared memory, hiding the penalty of cold cache misses and coherence issues.

\begin{algorithm}[t]
\scriptsize
\DontPrintSemicolon
i := getWarpToBeScheduled()\;
InstNo := getNumInstructions()\;
ActiveWarpNo := getNumActiveWarp()\;
\uIf{Warp(i).V == 0 \textbf{and} end of low cut-off epoch}{
	\tcc{Warp(i) is stalled}
	k := Pair\_List[i][1]\;
	$IRS_k$ := $^{VTAHit[k]}/_{InstNo/ActiveWarpNo}$\;
    \uIf{$IRS_k$ \textgreater low-cutoff \textbf{and} Warp(k) needs executing}{
        \textbf{continue}\;
    }
    \uElse{
        Warp(i).V := 1\;
        Pair\_List[i][1] := -1 \tcp{cleared}           }                         }
\uElseIf{Warp(i).I == 1 \textbf{and} end of low cut-off epoch}{
	\tcc{Warp(i) redirects to access shared memory}
	k := Pair\_List[i][0]\;
	$IRS_k$ := $^{VTAHit[k]}/_{InstNo/ActiveWarpNo}$\;
    \uIf{$IRS_k$ \textgreater low-cutoff \textbf{and} Warp(k) needs executing}{
        \textbf{continue}\;
    }
    \uElse{
        Warp(i).I := 0\;
        Pair\_List[i][0] := -1 \tcp{toggling}           }                         }
\uIf{Warp(i).V == 1 \textbf{and} end of high cut-off epoch }{
	\tcc{Warp(i) is active}
	$IRS_i$ := $^{VTAHit[i]}/_{InstNo/ActiveWarpNo}$\;
	j := Interference\_List[i]\;
	\uIf{$IRS_i$ \textgreater high-cutoff \textbf{and} $j$ != $i$ }{
	\uIf{ Warp(j).I == 1}{
		Warp(j).V := 0\;
		Pair\_List[j][1] := i\;
	}
	\uElseIf{ Warp(j).I == 0}{
		Warp(j).I := 1\;
		Pair\_List[j][0] := i\;
	}

	}
}
\caption{\arch scheduling algorithm}
\label{algo:CIAO}
\end{algorithm}

\subsection{Putting It All Together}
\label{sec:pat}
\noindent
Algorithm~\ref{algo:CIAO} describes how \arch schedules warps. 
For every \texttt{low-cutoff} epoch, the warp at the front of the warp list (\eg, warp $i$), is examined 
to decide whether \arch redirects memory requests of warp $i$ back to L1D cache or reactivate warp $i$.
More specifically, \arch first checks the first or second field of the \textit{pair list} entry corresponding to warp $i$. 
Once \arch confirms that either \arch previously redirected memory requests of warp $i$ to shared memory or stalled warp $i$ because warp $i$ severely interfered with another warp (\eg, warp $k$),
it redirects the memory requests of warp $i$ back to L1D cache or reactivate warp $i$, 
unless the following two conditions are satisfied: 
(1) $IRS_k$ is still higher than \texttt{low-cutoff} and (2) warp $k$ has not completed its execution.

Every \texttt{high-cutoff} epoch, \arch examines $IRS_i$.
If warp $i$ is in the active warp list and $IRS_i$ is higher than \texttt{high-cutoff}, 
\arch looks up the \textit{interference} list to determine which warp has most severely interfered with warp $i$. 
Once \arch determines the most interfering warp (\eg, warp $j$) for warp $i$, 
\arch checks whether it has redirected memory requests of warp $j$ to shared memory or stalled warp $j$. 
If \arch sees that warp $j$ has still sent memory requests to L1D cache, it isolates warp $j$, redirects memory requests of warp $j$ to shared memory,
and records warp $i$ in the first field of the pair list entry corresponding to warp $j$ to indicate that warp $i$ has triggered to redirect memory requests of warp $j$. 
If \arch has already redirected memory requests of warp $j$, then 
\arch starts to stall warp $j$ and records warp $i$ in second field of the pair list entry corresponding to warp $j$.
This record can be referenced when \arch decides to reactivate warp $i$ in future.

\section{Evaluation}
\label{sec:result}
\begin{table}
\begin{center}
\small
\begin{tabular}{|p{2.2cm}|p{5.5cm}|}
\hline
\# of SMs/threads            & 15, max 1536 per SM                                                                                                        \\ \hline
L1D cache             & 16KB w/ 128B lines, 4 ways, write no-allocate, local write-back, global write-through, \newedit{1-cycle latency} and LRU								      \\ \hline
\newedit{Shared memory}            & \newedit{48KB, 1-cycle latency and 32 banks} \\ \hline
L2 cache              & 768KB w/ 128B lines, 8 ways, write allocation, write-back and LRU			                     \\ \hline
DRAM                  & GDDR5 w/ 16 banks, tCL=12, tRCD=12, and tRAS=28               \\ \hline
Victim tag array      & 8 tags per set, 48 sets, and FIFO                             \\ \hline
\end{tabular}
\end{center}
\caption{\label{tab:config} GPGPU-Sim configuration.}
\end{table}


\begin{table}
\centering
\small
\begin{tabular}{|m{2.2cm}|m{0.5cm}|m{0.8cm}|m{0.6cm}|m{0.6cm}|m{0.4cm}|m{0.5cm}|}
\hline
Benchmark                                    & APKI & Input & $\rm N_{wrp}$ & $\rm F_{smem}$ &  Bar.& Class \\ \hline
\texttt{ATAX} \cite{pouchet2012polybench}    &    64 &  64MB &             2 &           0\% &  N   & LWS   \\ \hline
\texttt{BICG} \cite{pouchet2012polybench}    &    64 &  64MB &             2 &           0\% &  N   & LWS   \\ \hline
\texttt{MVT} \cite{pouchet2012polybench}     &    64 &  64MB &             2 &           0\% &  N   & LWS   \\ \hline
\texttt{KMN} \cite{he2008mars}               &    46 & 168KB &             4 &           1\% &  Y   & LWS   \\ \hline
\texttt{Kmeans} \cite{che2009rodinia}        &    85 & 101MB &             2 &           0\% &  Y   & LWS   \\ \hline \hline
\texttt{GESUMMV} \cite{pouchet2012polybench} &   136 & 128MB &             2 &           0\% &  N   & SWS   \\ \hline
\texttt{SYR2K} \cite{pouchet2012polybench}   &   108 &  48MB &             6 &           0\% &  N   & SWS   \\ \hline
\texttt{SYRK} \cite{pouchet2012polybench}    &    94 & 512KB &             6 &           0\% &  N   & SWS   \\ \hline
\texttt{II} \cite{he2008mars}                &    75 &  28MB &             4 &           0\% &  Y   & SWS   \\ \hline
\texttt{PVC} \cite{he2008mars}               &    64 &  13MB &            48 &          33\% &  Y   & SWS   \\ \hline
\texttt{SS} \cite{he2008mars}                &    34 &  23MB &            48 &          50\% &  Y   & SWS   \\ \hline
\texttt{SM} \cite{he2008mars}                &   140 &   1MB &            48 &           1\% &  Y   & SWS   \\ \hline
\texttt{WC} \cite{he2008mars}                &    19 &  88KB &            48 &           1\% &  Y   & SWS   \\ \hline \hline
\texttt{Gaussian} \cite{che2009rodinia}      &    18 & 339KB &            48 &           0\% &  N   & CI   \\ \hline
\texttt{2DCONV} \cite{pouchet2012polybench}  &     9 &  64MB &            36 &           0\% &  N   & CI   \\ \hline
\texttt{CORR} \cite{pouchet2012polybench}    &    10 &   2MB &            48 &           0\% &  N   & CI   \\ \hline 
\texttt{Backprop} \cite{che2009rodinia}      &     3 &   5MB &            36 &          13\% &  Y   & CI   \\ \hline
\texttt{Hotspot} \cite{che2009rodinia}       &     1 &   2MB &            48 &          19\% &  Y   & CI   \\ \hline
\texttt{Lud} \cite{che2009rodinia}           &     2 &  25KB &            38 &          50\% &  Y   & CI   \\ \hline
\texttt{NN} \cite{che2009rodinia}            &     8 & 334KB &            48 &           0\% &  N   & CI   \\ \hline
\texttt{NW} \cite{che2009rodinia}            &     5 &  32MB &            48 &          35\% &  Y   & CI   \\ \hline 
\end{tabular}
\caption{Benchmark characteristics. $\rm N_{wrp}$ and $\rm F_{smem}$ denote the number of active warps achieving the highest performance for \texttt{Best-SWL} and the fraction of shared memory used by application running on baseline GPU.}
\label{tab:workload_charac}
\end{table}

\begin{figure*}
\centering
\subfloat[IPC comparison.]{\label{fig:IPC_fig}\rotatebox{0}
{\includegraphics[width=0.8\linewidth]{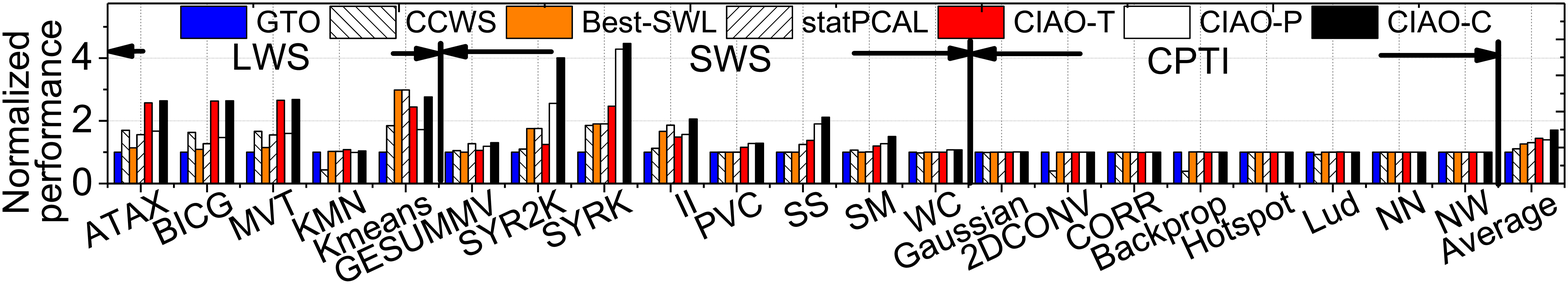}}}
\subfloat[Shared mem utilization.]{\label{fig:shmutil_fig}\rotatebox{0}
{\includegraphics[width=0.18\linewidth]{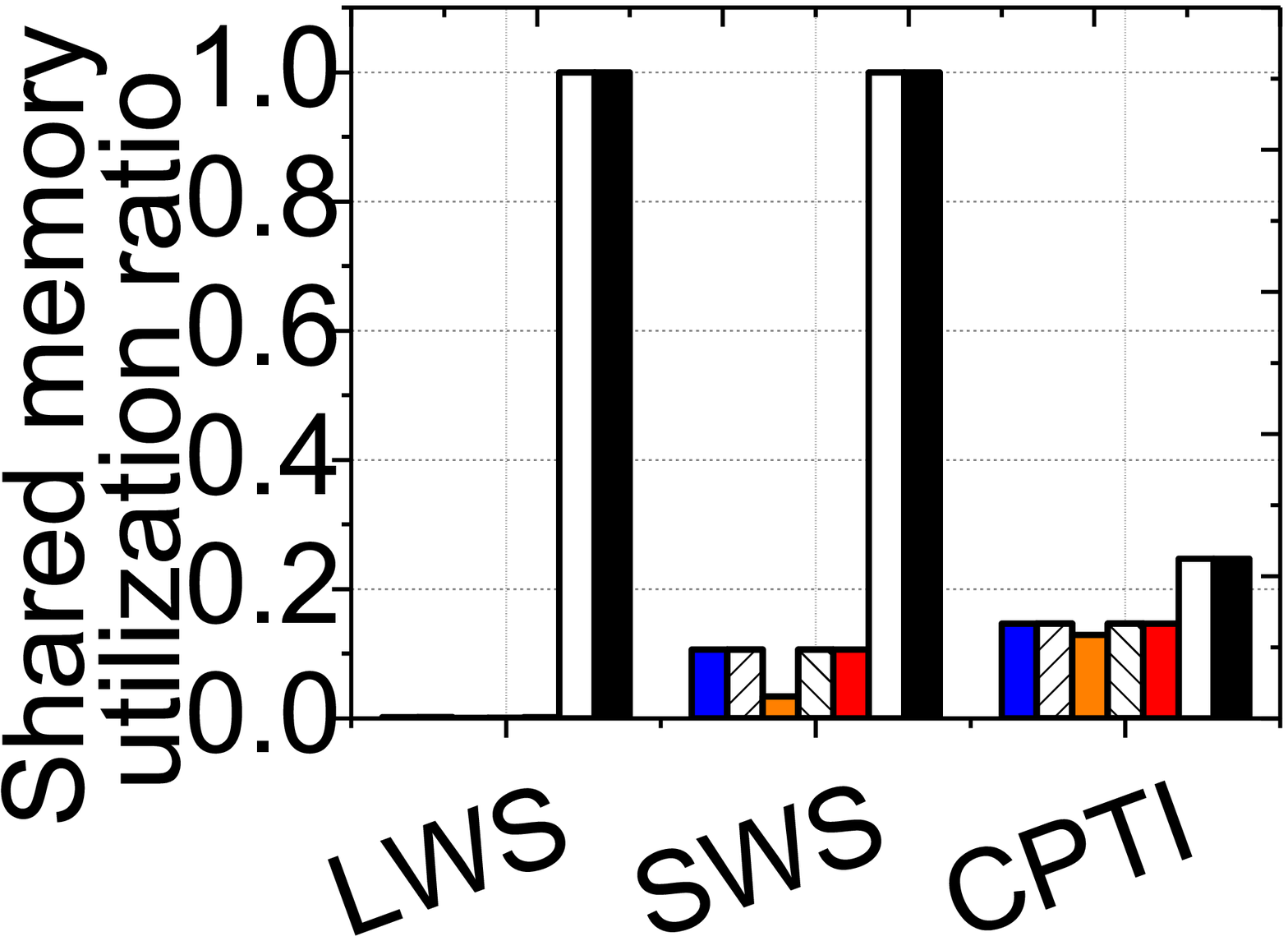}}}
\caption{Performance analysis; IPC values of each warp scheduler are normalized to those of \texttt{GTO}. 
}
\label{fig:overall_ipc}
\end{figure*}

\subsection{Methodology}
\label{sec:method} 
\noindent\textbf{GPU architecture.} 
We use GPGPU-Sim 3.2.2~\cite{aaamodt2012gpgpu} and configure it to model a GPU similar to NVIDIA GTX 480;
see Table \ref{tab:config} for the detailed GPGPU-Sim configuration parameters~\cite{nvidia2009nvidia}.
Besides, we enhance the baseline L1D and L2 caches with a XOR-based set index hashing technique~\cite{nugteren2014detailed}, making it close to the real GPU device's configuration. 
Subsequently, we implement seven different warp schedulers: 
(1) \texttt{GTO} (GTO scheduler with set-index hashing \cite{nugteren2014detailed});
(2) \texttt{CCWS};
(3) \texttt{Best-SWL} (best static wavefront limiting);
(4) \texttt{statPCAL} (representative implementation of bypass scheme\cite{li2015priority} that performs similar or better than \cite{li2015locality,tian2015adaptive});
(5) \texttt{CIAO-P} (\texttt{CIAO} with only redirecting memory requests of interfering warp to shared memory); 
(6) \texttt{CIAO-T} (\texttt{CIAO} with only selective warp throttling); and
(7) \texttt{CIAO-C} (\texttt{CIAO} with both \texttt{CIAO-T} and \texttt{CIAO-P}).
Note that \texttt{CCWS}, \texttt{Best-SWL}, and \texttt{CIAO-P/T/C} leverage \texttt{GTO} to decide the order of execution of warps. 
\texttt{CCWS} and \texttt{CIAO-T/C} stall a varying number of warps depending on memory access characteristics monitored at runtime.
In contrast, \texttt{Best-SWL} stalls a fixed number of warps throughout execution of a benchmark; we profile each benchmark to determine the number of stalled warps giving the highest performance for each benchmark; see column $\rm N_{wrp}$ in Table~\ref{tab:workload_charac}.

\noindent \textbf{Benchmarks.} We evaluate a large collection of benchmarks from \texttt{PolyBench}~\cite{grauer2012auto}, \texttt{Mars}~\cite{he2008mars} and \texttt{Rodinia}~\cite{che2009rodinia}
which are categorized into three classes: 
(1) large-working set (LWS), (2) small-working set (SWS), and (3) compute-intensive (CI). 
Table~\ref{tab:workload_charac} tabulates chosen benchmarks and their characteristics. 


\begin{figure*}
\centering
\subfloat[IPC]{\label{fig:ATAX_back_IPC_fig}\rotatebox{0}{\includegraphics[width=0.33\linewidth]{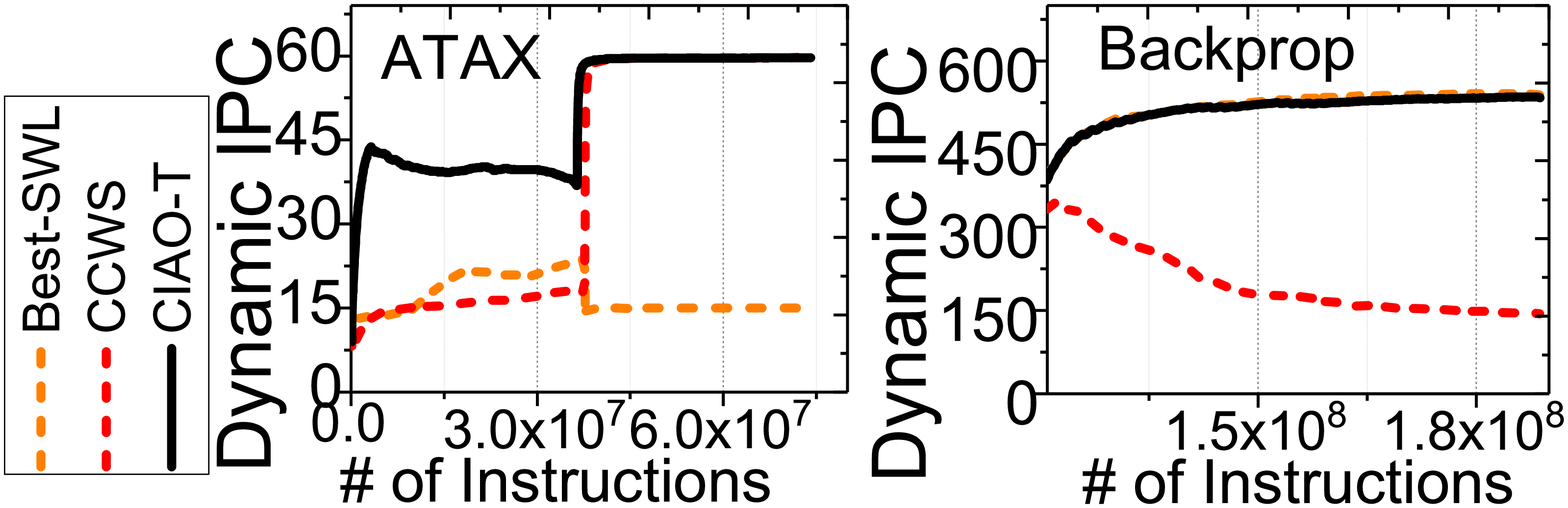}}}
\subfloat[Number of active warps]{\label{fig:ATAX_back_AW_fig}\rotatebox{0}{\includegraphics[width=0.33\linewidth]{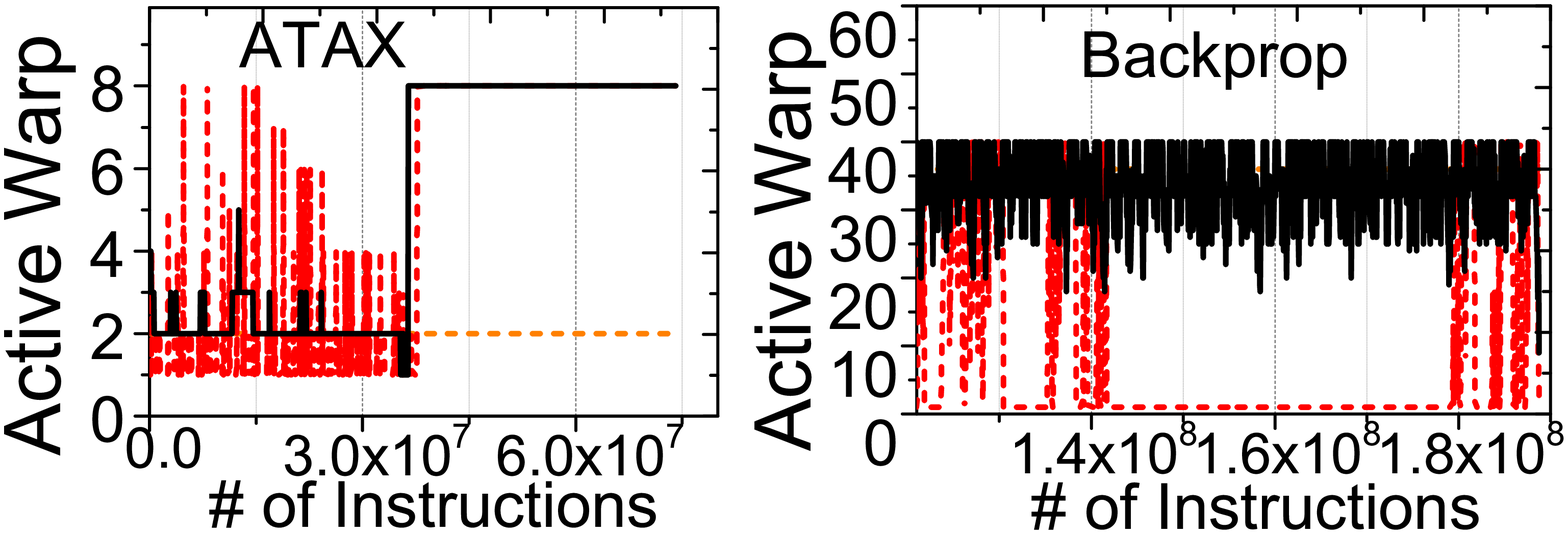}}}
\subfloat[Cache interference]{\label{fig:ATAX_back_interf_fig}\rotatebox{0}{\includegraphics[width=0.33\linewidth]{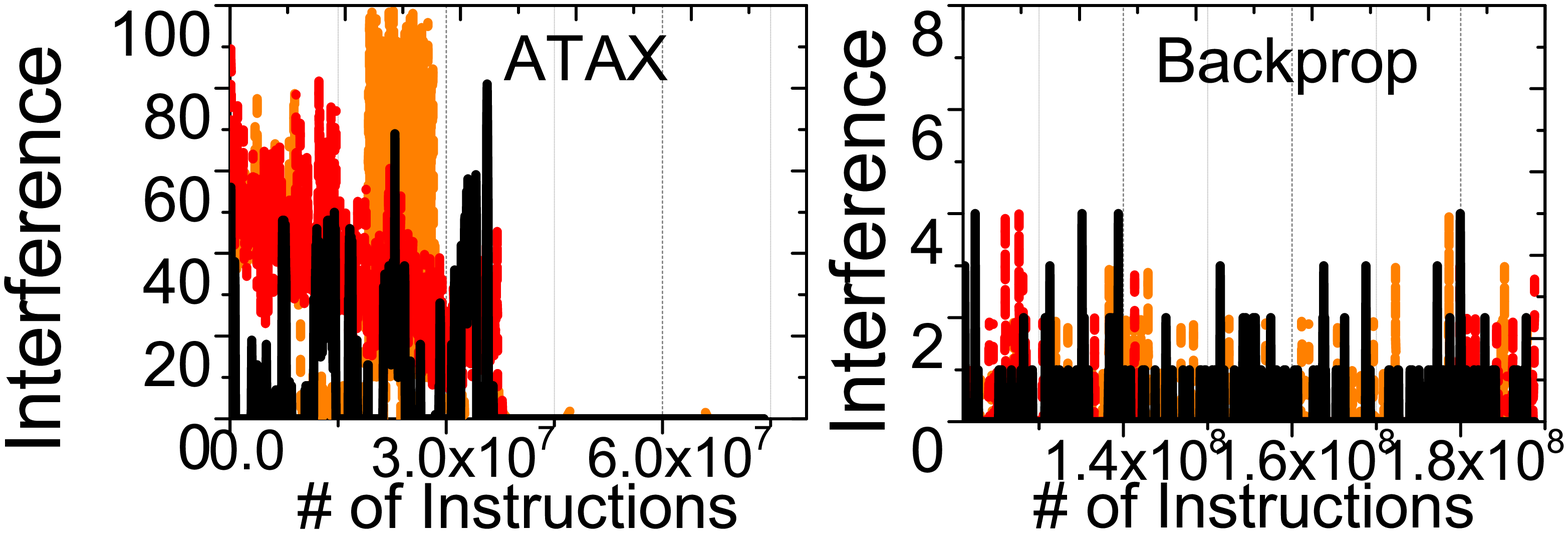}}}
\caption{Comparison between \texttt{Best-SWL}, \texttt{CCWS} and \texttt{CIAO-T} over time: \texttt{ATAX} and \texttt{Backprop}}
\label{fig:IPCtrace_ATAX_BACK}
\end{figure*}

\begin{figure*}
\centering
\subfloat[IPC]{\label{fig:SYRK_KMN_IPC}\rotatebox{0}{\includegraphics[width=0.33\linewidth]{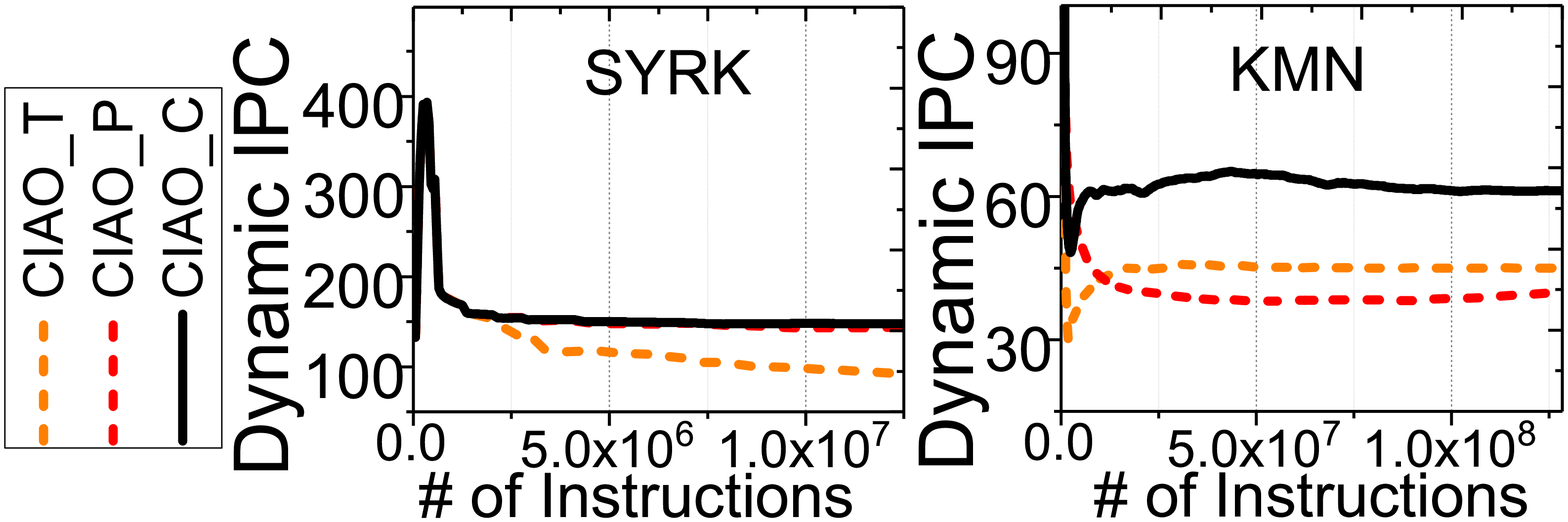}}}
\subfloat[Number of active warps]{\label{fig:SYRK_KMN_activewarp}\rotatebox{0}{\includegraphics[width=0.33\linewidth]{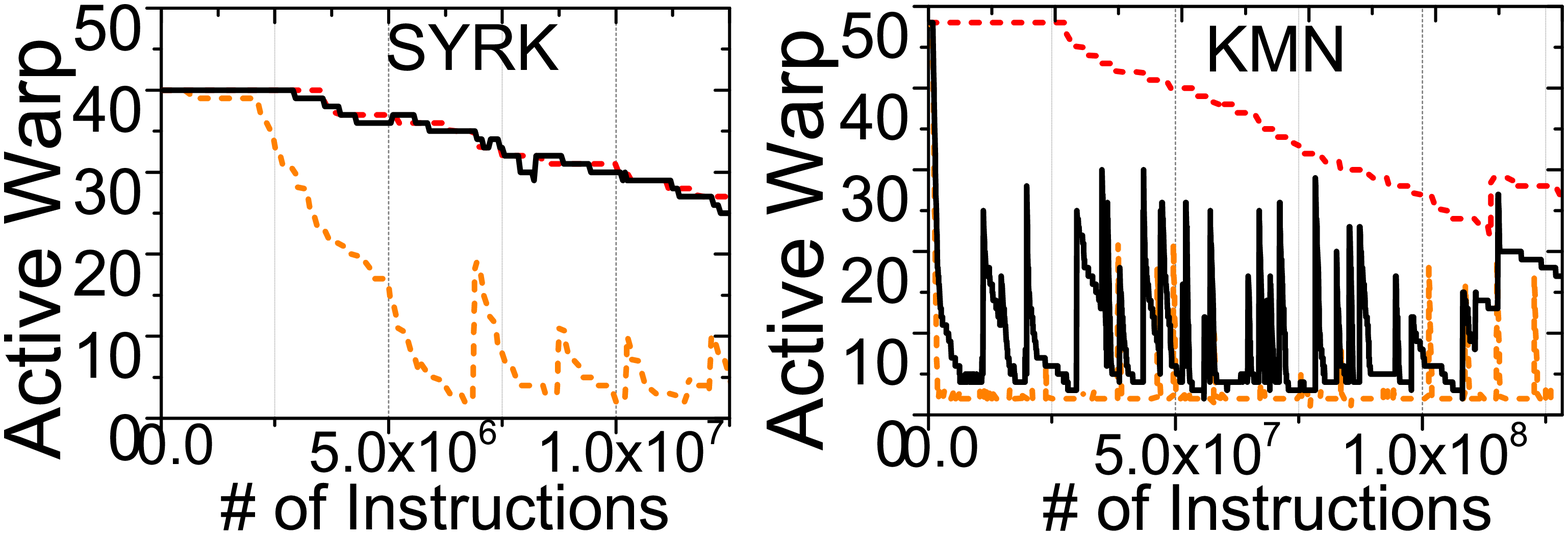}}}
\subfloat[Cache interference]{\label{fig:SYRK_KMN_interference}\rotatebox{0}{\includegraphics[width=0.33\linewidth]{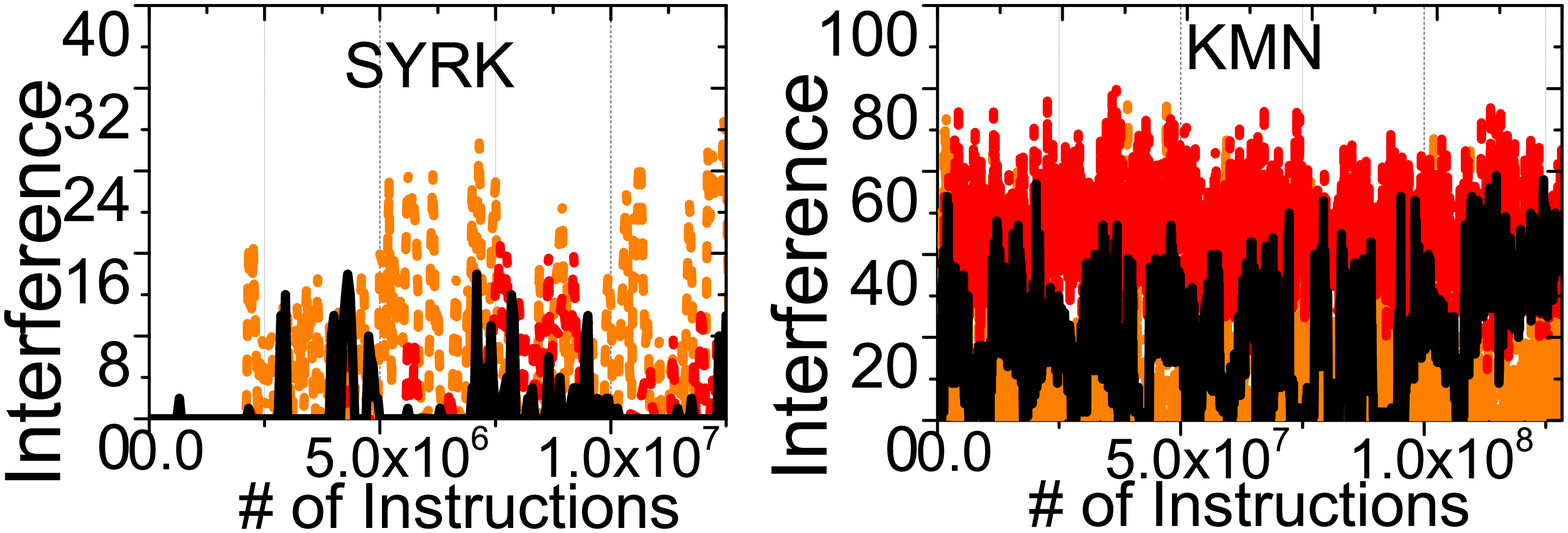}}}
\caption{Comparison of \texttt{CIAO-T}, \texttt{CIAO-P} and \texttt{CIAO-C} over time: \texttt{SYRK} and \texttt{KMN}.}
\label{fig:IPCtrace_SYRK_seperate}
\end{figure*}

\ignore{
\begin{figure*}
\centering
\subfloat[IPC]{\label{fig:ATAX_back_IPC_fig}\rotatebox{0}{\includegraphics[width=1\linewidth]{figs/ATAX_back_SYRK_KMN_IPC_fig}}}
\subfloat[Number of active warps]{\label{fig:ATAX_back_AW_fig}\rotatebox{0}{\includegraphics[width=1\linewidth]{figs/ATAX_back_SYRK_KMN_AW_fig}}}
\subfloat[Cache Interference]{\label{fig:ATAX_back_interf_fig}\rotatebox{0}{\includegraphics[width=1\linewidth]{figs/ATAX_back_SYRK_KMN_interf_fig}}}
\caption{Performance analysis of \texttt{ATAX}, \texttt{Backprop}, \texttt{SYRK}, and \texttt{KMN}.
}
\label{fig:IPCtrace}
\end{figure*}
}

\subsection{Performance Analysis}
\label{sec:analy}
\noindent
Figure~\ref{fig:overall_ipc}
plots the IPC values with the seven warp schedulers and the \textbf{geometric-mean} IPC values of three benchmark classes (LWS, SWS, and CI), respectively,
normalized to those with \texttt{GTO}. 
Overall, \texttt{CCWS}, \texttt{Best-SWL}, \texttt{statPCAL}, and \texttt{CIAO-C} provide 2\%, 16\%, 24\% and 56\% higher performance than \texttt{GTO}, respectively.

\texttt{GTO} performs worst among all evaluated schedulers, because, it shuffles only the order of executed warps and does not notably reduce cache thrashing caused by many active warps accessing small L1D cache. 
In contrast, \texttt{Best-SWL} outperforms \texttt{GTO} as it throttles some warps, reducing the number of memory accesses to small L1D and thus cache thrashing.  
Nonetheless, as \texttt{Best-SWL} must decide the number of throttled warps before execution of a given application, 
it cannot effectively capture the optimal number of throttled warps varying within an application compared to warp schedulers that dynamically throttle the number of executed warps such as \texttt{CCWS} and \texttt{CIAO}.
For example, as \texttt{ATAX} exhibits very dynamic cache access patterns at runtime, 
\texttt{CCWS} outperforms \texttt{Best-SWL} by 49\%. 
Note that \texttt{CCWS} gives notably lower performance than \texttt{Best-SWL} 
especially for CI benchmarks; considerably affecting its performance.
That is because running more active warps achieves higher performance for CI benchmarks, whereas \texttt{CCWS} unnecessarily stalls some active warps to give a higher priority to a few warps exhibiting high data locality.
\texttt{statPCAL} gives up to 37\% higher performance than \texttt{Best-SWL} by up to 37\% because \texttt{statPCAL} offers higher TLP.
Specifically, when \texttt{statPCAL} detects under-utilization of L2 and/or main memory bandwidth, it activates throttled warps and makes these warp directly access the underlying memory (\ie, bypassing L1D cache).
Due to the long access latency and limited bandwidth of underlying memory, however,
\texttt{statPCAL} cannot significantly improve performance of LWS and SWS workloads such as \texttt{KMN}, \texttt{SYRK}, etc. 

\texttt{CIAO-T} provides 32\% and 34\% higher performance than \texttt{CCWS} and \texttt{GTO}, respectively. 
Furthermore, \texttt{CIAO-T} offers 22\% higher performance than \texttt{Best-SWL} for every benchmark except for \texttt{SYR2K}, \texttt{II}, and \texttt{KMN} exhibiting static cache access patterns at runtime.
Both \texttt{CIAO-T} and \texttt{CCWS} dynamically stall some active warps at runtime, but
our evaluation shows that it is often more effective to throttle the warps that considerably interfere with other warps than the warps with low potential of data locality
as \texttt{CCWS} does.
Furthermore, for CI benchmarks, \texttt{CIAO-T} offers as high performance as \texttt{GTO} in contrast to \texttt{CCWS}; 
refer to our earlier comparison between \texttt{GTO} and \texttt{CCWS} for CI benchmarks.

\texttt{CIAO-P}
gives 34\% higher performance than \texttt{GTO}. 
\newedit{We observe that \texttt{CIAO-P} offers the highest TLP among all seven warp schedulers, entailing 28\% higher performance than \texttt{CIAO-T} for SWS class benchmarks. This is because \texttt{CIAO-P} fully utilizes the unused space of shared memory (cf. Figure~\ref{fig:shmutil_fig}). }
Nonetheless, its benefits can be limited for LWS class benchmarks in which the redirected memory requests of interfering warps are often too intensive and thus thrash the shared memory as well.
In such a case, \texttt{CIAO-T} can perform better than \texttt{CIAO-P}, giving 48\% and 66\% higher performance than \texttt{CIAO-P} and \texttt{CCWS}, respectively,  as shown in Figure \ref{fig:IPC_fig}. 
Lastly, \texttt{CIAO-C}, which synergistically integrates \texttt{CIAO-T} and \texttt{CIAO-P}, provides 56\%, 54\%, 17\% and 16\% higher performance than \texttt{GTO}, \texttt{CCWS}, \texttt{CIAO-T}, and \texttt{CIAO-P}, respectively. 

\begin{figure}[b]
\centering
\subfloat[Various epoches.]{\label{fig:epoch}\rotatebox{0}{\includegraphics[width=0.48\linewidth]{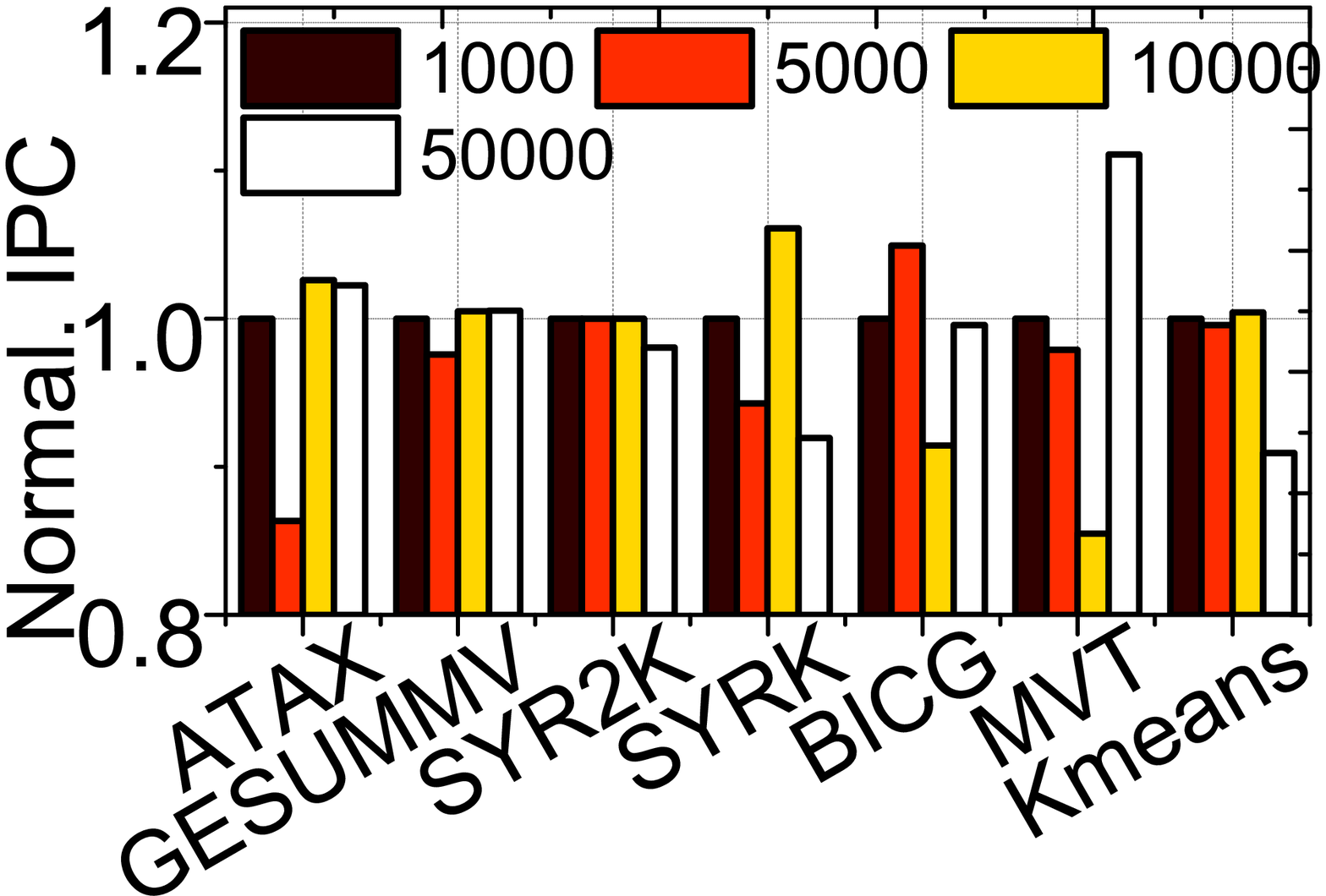}}}
\hspace{2pt}
\subfloat[Vairous high cut-off lines.]{\label{fig:highcutoff}\rotatebox{0}{\includegraphics[width=0.48\linewidth]{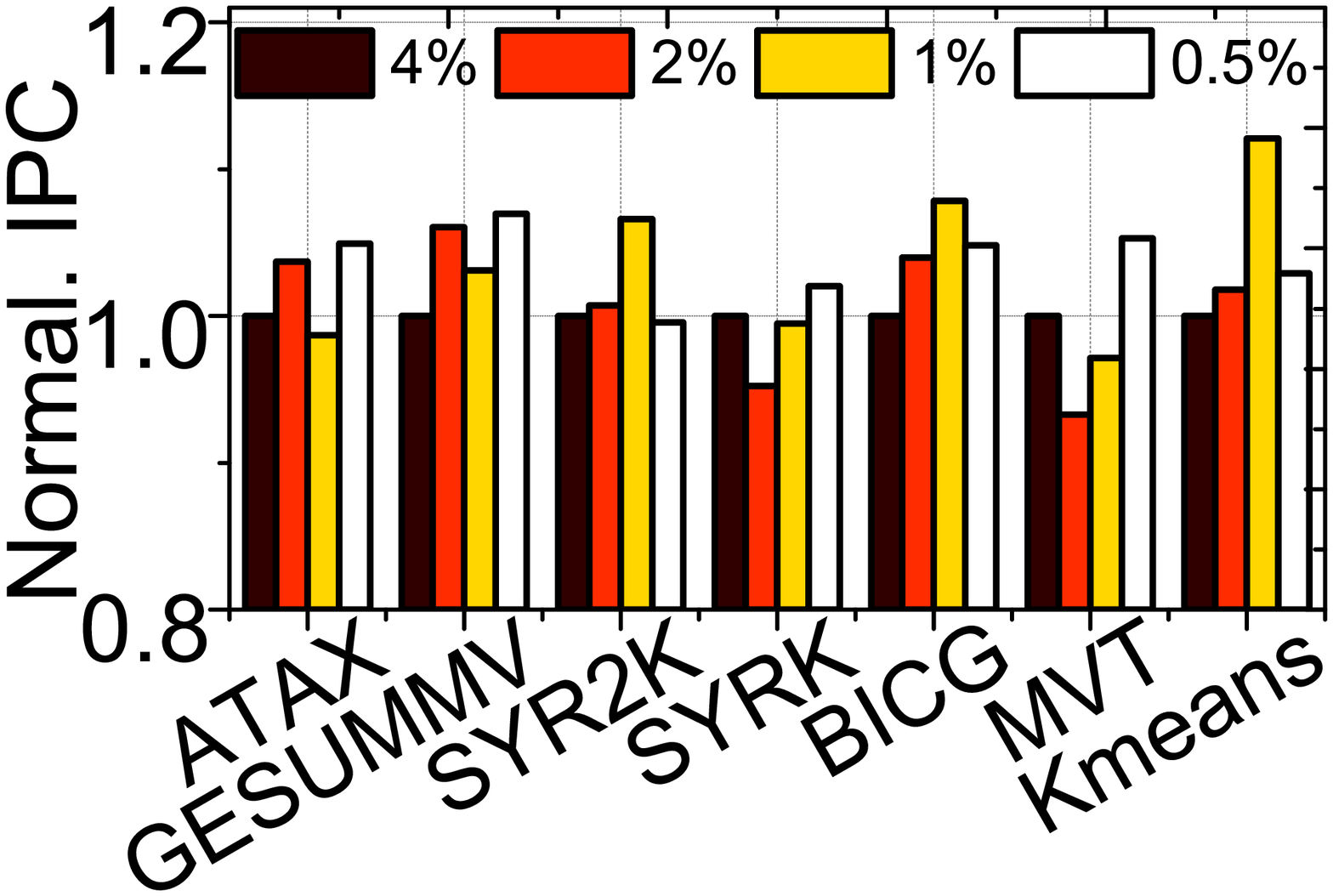}}}
\caption{Sensitivity analysis.
}
\label{fig:sensi_scheduler}
\end{figure}

\begin{figure*}
\centering
\subfloat[IPC comparison of varying L1D cache configurations.]{\label{fig:IPC_fig_sens}\rotatebox{0}
{\includegraphics[width=0.49\linewidth]{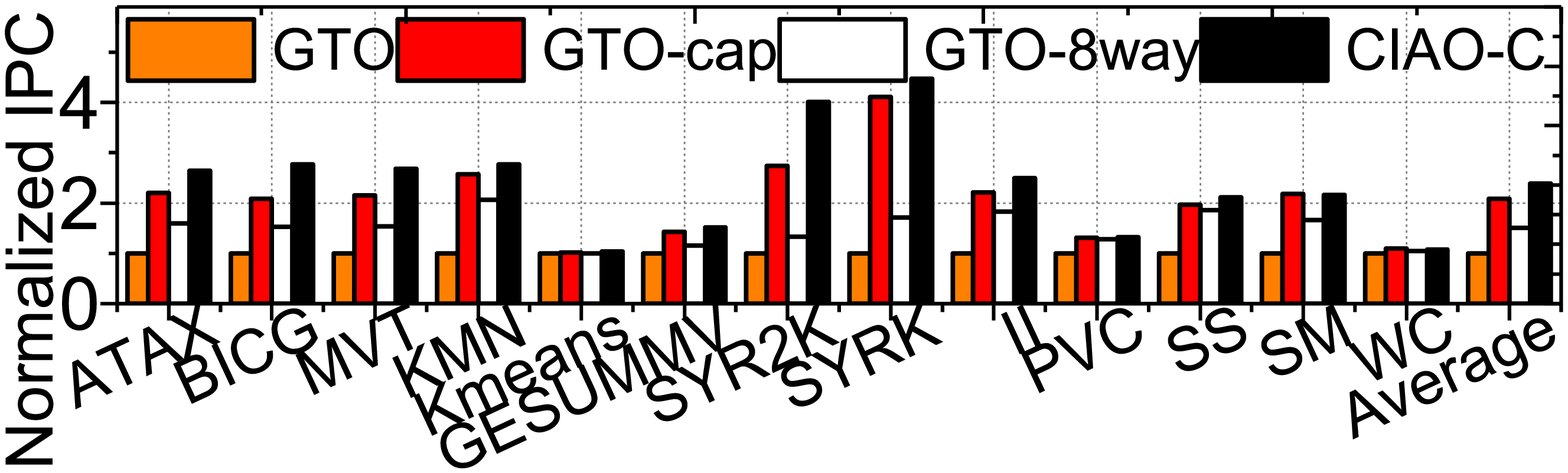}}}
\subfloat[IPC comparison of vayring DRAM bandwidths.]{\label{fig:IPC_fig_sens1}\rotatebox{0}
{\includegraphics[width=0.49\linewidth]{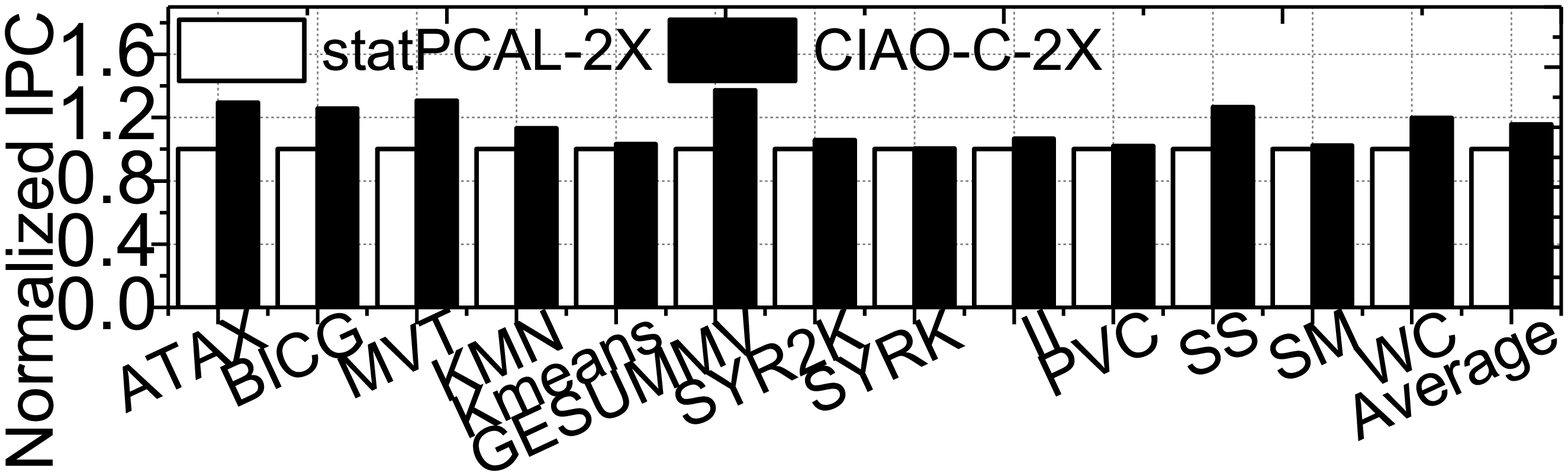}}}
\caption{IPC of different L1D cache and DRAM configurations.  
}
\label{fig:sensi1_cache}
\end{figure*}

\subsection{Effectiveness of Interference Awareness}
\noindent 
Figure~\ref{fig:IPCtrace_ATAX_BACK}
shows the IPC, the number of active warps, and cache interference over time of \texttt{ATAX} as a representative application that exhibits distinct execution phases in a single kernel execution. 
For example, \texttt{ATAX} exhibits two distinct execution phases.
The first phase comprised of the first 40-million instructions is very memory-intensive, whereas the second phase is very compute-intensive.
Figure \ref{fig:ATAX_back_IPC_fig} shows that \texttt{CIAO-T} outperforms \texttt{CCWS} and \texttt{Best-SWL} for the first 40-million instructions executed. 
\texttt{CIAO-T} exhibits higher performance during this phase because \texttt{CIAO-T} more effectively reduces cache interference by throttling severely interfering warps, as shown in Figure \ref{fig:ATAX_back_interf_fig}.
After the first phase, \texttt{ATAX} starts a compute-intensive phase, performing the computation by fully exploiting data locality on the GPU caches. 
As \texttt{Best-SWL} cannot capture this dynamics at runtime, it executes only 2 warps for the second phase execution of \emph{ATAX}. 
In contrast, \texttt{CCWS} and \texttt{CIAO-C} dynamically reduce the number of stalled warps as they observe fewer cache misses and less cache interference, 
giving 4$\times$ higher geometric-mean performance than \texttt{Best-SWL}. 


We choose \texttt{Backprop} as a representative application that is very compute-intensive but also experiences many cache misses.
Figure \ref{fig:IPCtrace_ATAX_BACK} shows the performance change of \emph{Backprop} over time.
\texttt{Best-SWL} and \texttt{CIAO-T} provide 500 IPC on average. 
However, \texttt{CCWS} notably degrades the performance, ranging from 320 to 150 IPC because \texttt{CCWS} ends up giving a higher priority to warps with higher data locality and stalling more than 40 warps (or significantly reducing TLP).
In contrast, \texttt{CIAO-T}, which offers performance similar to \texttt{Best-SWL}, more selectively throttles warps than \texttt{CCWS} (i.e., only 10$\sim$20 most interfering warps), better preserving TLP.

\subsection{Sensitivity to Working Set Size}
\label{sec:tsa}
\noindent \textbf{Small-working set.}
Figure~\ref{fig:IPCtrace_SYRK_seperate} shows the performance of three \texttt{CIAO} schemes for \texttt{SYRK} over time.
\texttt{SYRK} is a representative application with SWS. 
Specifically, 
Figure~\ref{fig:IPCtrace_SYRK_seperate}
illustrates IPC, the number of active warps, and the number of cache conflicts over time of texttt{SYRK} over time with three \texttt{CIAO} schemes. 
As shown in Figure \ref{fig:SYRK_KMN_IPC}, \texttt{CIAO-P} offers higher IPC than \texttt{CIAO-T} overall. 
This is because, it can secure higher TLP (\cf Figure \ref{fig:SYRK_KMN_activewarp}), whereas \texttt{CIAO-T} alone hurts TLP by throttling many active warps. 
Using the unused shared memory space, \texttt{CIAO-P} can effectively reduce cache interference without sacrificing TLP in contrast to \texttt{CIAO-T}. 
As expected, \texttt{CIAO-C} selectively stalls very few warps.

\noindent \textbf{Large-working set.}
Figure~\ref{fig:IPCtrace_SYRK_seperate} also depicts the performance of three \texttt{CIAO} schemes \texttt{KMN}, a representative application with LWS. 
As shown in Figure~\ref{fig:SYRK_KMN_IPC}, \texttt{CIAO-T} provides 50\% higher IPC than \texttt{CIAO-P}, and
\texttt{CIAO-C} always achieves the highest performance during the entire execution period amongst all three schemes. 
This is because, as shown in Figure \ref{fig:SYRK_KMN_interference}, \texttt{CIAO-P} still suffers from severe shared memory interference 
as the amount of data requested by the partitioned warps exceeds the amount that shared memory can efficiently accommodate. 
In contrast, \texttt{CIAO-C} can better utilize shared memory by selectively throttling only the warps that cause severe interference. 

\ignore{
\begin{figure*}
\centering
\subfloat[IPC comparison]{\label{fig:IPC_fig_16112}\rotatebox{0}
{\includegraphics[width=0.8\linewidth]{figs/IPC_fig_16112.eps}}}
\subfloat[Geo-mean IPC]{\label{fig:IPC_fig2_16112}\rotatebox{0}
{\includegraphics[width=0.18\linewidth]{figs/IPC_fig2_16112.eps}}}
\caption{IPC of 16KB L1D cache and 112KB shared mem.  
}
\label{fig:sensi_cache}
\end{figure*}
}

\ignore{the highest reduction in L2 miss rate comes from LRR + CIAO -- compared to LRR and LRR + \texttt{CCWS} by 52.7\% and 48.9\%, respectively. This dramatic decrease in miss rate results from a combined effect of the active warp number throttling and the consideration of cache interference upon warp scheduling. On the other hand,  \texttt{GTO} + CIAO reduces the L2 miss rate by 28.3\% and 13.8\% over \texttt{GTO} and \texttt{GTO} + \texttt{CCWS}, respectively. Due to the high number of active warps, even restricting to the oldest warps, \texttt{GTO} still allows too much data to be contained in L2 cache. Even though, \texttt{CCWS} can further alleviate the L2-level cache interference by strictly limiting the active warp number, warps with high potential of data locality, which are prioritized in \texttt{CCWS}, can still contend with each other. The scheduling policy of CIAO successfully exploits this critical observation regarding the potential for further reduction in L2-level interference.
}

\ignore{
\begin{figure}
\centering
\includegraphics[width=1\linewidth]{figs/L1Dynenergy_fig.eps}
\caption{L1 total energy analysis.  
}
\label{fig:L1Dynenergy_fig}
\end{figure}

\begin{figure}
\centering
\includegraphics[width=1\linewidth]{figs/L2dynenergy_fig.eps}
\caption{L2 total energy analysis. 
}
\label{fig:L2dynenergy_fig}
\end{figure}

\begin{figure}
\centering
\includegraphics[width=1\linewidth]{figs/DRAMdynenergy_fig.eps}
\caption{DRAM dynamic energy analysis. 
}
\label{fig:DRAMdynenergy_fig}
\end{figure}
}

\subsection{Sensitivity Study}
\label{subsubsec:sensi}
\noindent \textbf{Epoch value.} 
Figure~\ref{fig:epoch} shows the effect of varying \texttt{high-cutoff} epoch values on the IPC for all the memory-intensive workloads. 
As we increase the epoch from 1K to 50K instructions, the change in IPC is within 15\%. 
Note that different workloads can achieve best performance with different epoch values. 
That is because epoch determines the frequency of checking cache interference for \texttt{CIAO}.
A shorter epoch provides fast response to cache interference, while a longer epoch can more accurately detect the warp causing most interference. 
Taking this trade-off into account, we choose 5K instructions as \texttt{high-cutoff} epoch value. 
An adaptive scheme can be future work.

\noindent \textbf{High-cutoff threshold.} 
Figure~\ref{fig:highcutoff} depicts performance corresponding to different \texttt{high-cutoff} thresholds, 
where the \texttt{low-cutoff} threshold is fixed to half of it. 
All benchmarks show steady performance within 5\% change during the entire execution period.
This is because our \texttt{CIAO} throttles the active warps causing most interference, which can easily exceed the current thresholds we set. 1\% is chosen in the paper. 


\newedit{
\noindent \textbf{L1D cache/DRAM configurations.} Figure~\ref{fig:sensi1_cache} illustrates the performance of LWS and SWS workloads by configuring various L1D cache/DRAM design parameters: 
(1) \texttt{GTO};
(2) \texttt{GTO-cap} (\texttt{GTO} but increase L1D cache capacity to 48 KB and reduce shared memory size to 16 KB);
(3) \texttt{GTO-8way} (\texttt{GTO} but increase L1D cache associativity to 8 way);
(4) \texttt{statPCAL-2X} (\texttt{statPCAL} but double DRAM bandwidth from 177 GB/s to 340 GB/s);
(5) \texttt{CIAO-C};
(6) \texttt{CIAO-C-2X} (\texttt{CIAO-C} but double DRAM bandwidth).
As shown in Figure \ref{fig:IPC_fig_sens}, while increasing L1D cache capacity (\texttt{GTO-cap}) and associativity (\texttt{GTO-8way}) can effectively improve the overall performance by 108\% and 51\% compared to \texttt{GTO}, \texttt{CIAO-C} still outperforms \texttt{GTO-cap} and \texttt{GTO-8way} by 14\%, and 57\%, respectively. This is because, \texttt{GTO-cap} and \texttt{GTO-8way} cannot fully eliminate cache interference, as they cannot distinguish the requests between interfering and interfered warps and effectively isolate them. On the other hand, while \texttt{statPCAL-2X} can benefit from the increased DRAM bandwidth, bypassing requests to underlying DRAM still suffers from long DRAM delay as the latency of DRAM access is much longer than that of L1D cache access. Hence, as shown in Figure \ref{fig:IPC_fig_sens1}, \texttt{CIAO-C-2X} outperforms \texttt{statPCAL-2X} by 16\%, on average.
}

\ignore{
\noindent \textbf{Varying L1D cache sizes.} 
Figure~\ref{fig:sensi1_cache} 
illustrates the performance of all workloads for the various L1D cache configurations shown in Table~\ref{tab:config}. 
For 48KB L1D cache and 16KB shared memory, \texttt{CIAO-C} gives 29\%, 27\%, 13\%, and 12\% higher IPC than \texttt{GTO}, \texttt{CCWS}, \texttt{Best-SWL}, and \texttt{statPCAL}, respectively.
Since 48KB L1D cache is much larger than the default 16KB size, the performance of \texttt{GTO} improves greatly and leaves less room for improvement by \arch. 
}

\subsection{Overhead Analysis}
\noindent 
Implementing the interference detector, \texttt{CIAO} leverages the VTA structure originally proposed by \texttt{CCWS}~\cite{rogers2012cache}, but employs only 8 VTA entries for each warp (\ie, half of the VTA entries that \texttt{CCWS} uses).
Using CACTI 6.0~\cite{muralimanohar2009cacti}, we estimate that the area of one VTA structure is only 0.65 $mm^2$ for 15 SMs, which accounts for only 0.12\% of the total chip size of NVIDIA GTX480 (529 $mm^2$~\cite{geforce-gtx-480}).
In addition, \texttt{CIAO} uses 48 registers as VTA-hit counters (one for each warp). 
Since each VTA-hit counter resets at the start of each kernel, a 32-bit counter is sufficient to prevent its overflow. 
The interference and pair lists are implemented with SRAM arrays indexed by WIDs. 
Since the total number of active warps in a CTA does not exceed 64 (usually, 48 active warps in each SM), we configure the interference and pair lists with 64 entries. 
Each entry of the interference list requires 8 (= 6+2) bits to store one warp index and saturation counter value, while each entry of pair list requires 12 (=6 + 6) bits to store two warp indices.
Using CACTI 6.0, we estimate that the combined area of the VTA-hit counters, interference list, and pair lists is 549 $um^{2}$ per SM (8235 $um^{2}$ for 15 SMs).
On the other hand, Equation \ref{eq:irs} is implemented with a few adders, a shifter, and a comparator, which also requires very low cost (2112 gates). 
For our shared memory modification, the translation unit, multiplexer and MSHR only need 4500 gates and 64B storage per SM.
We also track the power consumption of new components employed in \texttt{CIAO} by leveraging GPUWattch~\cite{leng2013gpuwattch}, 
which reveals the average power is around 79mW. 
Overall, \texttt{CIAO} improves the performance by more than 50\% with a negligible area cost (less than 2\% of the total GTX480 chip area) and power consumption (only 0.3\% of GTX480 overall power).

\section{Discussion and Related Work}
\label{sec:relatedwork}
\ignore{
\noindent \textbf{Area overhead analysis.}
For the interference detector, \texttt{CIAO} leverages the VTA structure originally proposed by \texttt{CCWS}~\cite{rogers2012cache}, but employs only 8 VTA entries for each warp (i.e., half of what \texttt{CCWS} uses) 
based on our observation that a small number of warps (4 or 5) cause most of cache interference for most applications.
Based on CACTI 6.0~\cite{muralimanohar2009cacti} we estimate that the area of one VTA structure is only 0.65 $mm^2$ (XXX for 15 SMs) in a 22nm technology. 
That is, only 0.12\% of the total chip size of NVIDIA GTX480 (529 $mm^2$~\cite{}).
In addition, \texttt{CIAO} needs 48 32-bit registers for VTA-hit counters (one for each warp). 
Since each VTA-hit counter resets at the start of each kernel, one 32-bit register is sufficient to prevent a VTA-hit counter from overflowing. 
The interference and pair lists are SRAM register arrays indexed by WID. 
The interference list also needs to maintain the saturation counter value for each entry. 
Since the total number of active warps in a CTA does not exceed 64 (usually, 48 active warps in each SM) and 2 bits are required for saturation prediction, 
each entry of the interference and pair lists requires 8 (= 6+2) bits and 12 (=6 + 6) bits, respectively.
Based on CACTI 6.0~\cite{muralimanohar2009cacti}, we estimate that the combined area of the interference list and pair lists is 249.7 $um^{2}$ per SM (3746 $um^{2}$ for 15 SMs) in a 22nm technology.
Overall, \texttt{CIAO} improve the performance by more than 50\% at a negligible cost (less than 2\% of the total GPU chip area).
}
\ignore{
\noindent \textbf{Critical path analysis.} In our scheduler, the IRS evaluation may bring a burden to calculation. However, the actual implementation of Equation \ref{eq:irs} can be simplified as a few add, shift and comparison operations, which cost few cycles for the calculation. 
In addition, to minimize the data access delay to the interference list and pair list, we employ 8-bit and 12-bit data ports, which equal to the width of these two lists. Derived from CACTI6.0, such array access takes 0.23ns, which is much less than 1 GPU cache cycle. To make sure our modification does not impact critical path, the data locality is evaluated once in each epoch, which can hide the calculation latency.
}

\ignore{
\noindent \textbf{L2 optimization.} \remark{In the revised paper, L2 will not be discussed.} \newedit{Cache compression technology is widely accepted in last level cache design to efficiently expand the logical cache lines in one cache set, but keep the original cache block frame \cite{}. Since cache compression implementation does not require modifying L2 indexing mechanism to find out a target cache set, cache compression technique is orthogonal to our \texttt{CIAO}.}

\noindent \textbf{Multi-kernel execution.} \remark{This discussion is also about the correlation of L1-L2. It should be deleted.} \newedit{In our current work, we assume a GPU device executes a single-kernel in one time. Based on this assumption, we observe the correlation of L1D and L2 cache interference and propose an interference aware scheduling approach. However, the observation and solution derived from single-kernel execution cannot directly apply to multi-kernel execution. For example, multi-kernel execution may change the accuracy of IRS-calculation. In addition, different kernels running on different GPU cores can produce different access pattern, which may impact the L1-L2 correlation. The challenges of multi-kernel execution on \texttt{CIAO} will be addressed in our future work.}

\noindent \textbf{Barrier.} \remark{I will discuss barrier in shared memory.}\newedit{Prior work \cite{} \cite{} propose to balance the execution time of each warp, so that stagger warps will not degrade the performance in front of barrier or reconvergence point.
However, considering the negative impact of cache interference, we believe even with few barriers and reconvergence points, warp throttling approach \texttt{CIAO} can accelerate performance. If we do not stall the active warp that interferes with the other warps' cache accesses, all of them can suffer from the overhead imposed by long memory latency, which in turn can make overall latency even worse. In our approach, even though we stall active warps, all the other warps can reach the barrier/ reconvergence point with less interference. Also, the stalled warps can be reactivated as soon as possible. Our evaluation covers several workloads which contain barriers. From the evaluation results, \texttt{CIAO} does not degrade the performance compared to GTO because of barrier issue.
}
}

\noindent \textbf{Warps scheduling.} Several prior studies proposed to improve GPU performance by optimizing the warp scheduling methods.
The two-level warp scheduler \cite{narasiman2011improving} statically clusters active warps into several groups and executes the warp groups with different time intervals. This approach can alleviate memory traffic by reducing the number of memory accesses.
DYNCTA \cite{kayiran2013neither} and OWL \cite{jog2013owl} introduced data-locality aware schedulers that dynamically limit the number of active warps at runtime, which can address cache conflict problem. Similarly, DAWS \cite{rogers2013divergence} can predict data locality more accurately using runtime memory and branch divergence information. Unfortunately, all these schedulers do not acknowledge the cache interference exhibited by multiple active warps, are also correlated to data locality that they prioritize for warp scheduling. Further, as these schedulers throttle the active warps to address memory congestion issue, they may not able to secure high TLP. In contrast, our \texttt{CIAO} monitors cache interference and partitions/throttles few thread groups that exhibit cache thrashing and interference. This allows our scheduler to remove unnecessary memory accesses, maintain higher data locality, and secure high TLP.

\noindent \textbf{L1D cache and shared memory layout.} There exist some GPU models such as Pascal \cite{PASCAL} that simplify the complexity of on-chip memory design by splitting the L1D cache and shared memory into separate memory structures. \newedit{While prior work \cite{gebhart2012unifying}, which proposes to flexibly partition the unified on-chip memory by users, are not compatible with these GPU models, \arch well suits with their separate memory structure.} This is because our warp partition approach introduces modified MSHR and a new address translation unit to support independent data access in shared memory.

\noindent \textbf{Unbalanced cache interference.} 
As stated in section \ref{sec:interfere}, due to the limited cache capacity, 
warps of regular memory access patterns can generate cache interference. 
Such 
cache interference becomes more severe and non-uniform when warps experience irregular cache access pattern. 
One example is GPU execution of sparse matrix-vector multiplication (\texttt{SpMV}), which repeatedly accesses the large and irregular sparse matrix and vector. 
As typical implementations~\cite{su2012clspmv} refer sparse matrix/vector by an index array that contains indices of non-zero data, 
each warp can exhibit different irregularity based on given values of the index array.
\ignore{
\noindent \textbf{Overhead analysis.}
For the interference detector implementation, \texttt{CIAO} leverages the VTA structure originally proposed by \texttt{CCWS}~\cite{rogers2012cache}. 
Using CACTI 6.0~\cite{muralimanohar2009cacti}, we estimate that the area of one VTA structure is only 0.65 $mm^2$ for 15 SMs, which accounts for only 0.12\% of the total chip size of NVIDIA GTX480 (529 $mm^2$~\cite{geforce-gtx-480}).
In addition, \texttt{CIAO} uses 48 registers as VTA-hit counters (one for each warp). Since each VTA-hit counter resets at the start of each kernel, 32-bit width is sufficient to prevent a VTA-hit counter from binary overflow. 
The interference and pair lists are implemented as SRAM arrays, indexed by WID. Since the total number of active warps in a CTA does not exceed 64 (usually, 48 active warps in each SM), we configure the interference and pair lists with 64 entries. Each entry of the interference list requires 8 (= 6+2) bits to store one warp index and saturation counter value, while each entry of pair list requires 12 (=6 + 6) bits to store two warp indices.
Using CACTI 6.0, we estimate that the combined area of the VTA-hit counters, interference list, and pair lists is 549 $um^{2}$ per SM (8235 $um^{2}$ for 15 SMs).
On the other hand, Equation \ref{eq:irs} is implemented with a few adders, a shifter, and a comparator, which also requires very low cost (2112 gates). For our shared memory modification, the translation unit, MUX and MSHR only need 4500 gates and 64B storage per SM.
We also track the power consumption of new components employed in \texttt{CIAO} by leveraging GPUWattch \cite{leng2013gpuwattch}, which reveals the average power is around 79mW. Overall, \texttt{CIAO} improves the performance by more than 50\% with a negligible area cost (less than 2\% of the total GTX480 chip area) and power consumption (only 0.3\% of GTX480 overall power).
}
\ignore{
\noindent \textbf{Memory scheduling.} 
There are many prior studies that try to improve the throughput by increasing memory utilization. LAMAR \cite{rhu2013locality} consumes less memory bandwidth by managing the memory requests with a finer granularity. Specifically, it only fetches small amount of data requested by warps, instead of a whole cache block. Equalizer \cite{sethia2014equalizer} tunes the number of CTAs to improve performance and save energy by monitoring the amount of warps that wait for issuing memory requests. On the other hand, Mascar \cite{sethia2015mascar} can mitigate the memory access congestion issue by only allowing only single warp to access underlying memory if there is no available memory bandwidth. While all these approaches improve performance being aware of off-chip memory access patterns, none of them take in account the L1D and L2 cache access patterns, even though they significantly influence the runtime memory bandwidth.
MRPB \cite{jia2014mrpb} reorders the incoming memory requests to improve cache utilization. Specifically, it gathers the multiple memory requests that head to the same cache blocks, while bypassing all the memory requests that introduce cache pollution. Even though MRPB can increase the cache hit rates, its performance benefit unfortunately is limited in case where there are numerous irregular memory requests, which cause heavy off-chip traffic congestions due to its bypassing scheme. 
  Unlike these prior approaches, our \texttt{CIAO} can effectively mitigate the memory congestion problem by maximizing cache resource utilization. Our scheduler can preserve data frequently reused by the active warps, while arbitrating the cache interference observed in both L1 and L2 caches.
}
 

\section{Conclusion}
\label{sec:conclusion}
\noindent
In this work, 
we first shows that warps with high potential of data locality causes severe interference at L1D cache. 
Then we propose to (1) redirects memory requests of such warps to unused shared memory space and (2)
throttle such warps only when (1) still causes severe interference at shared memory;
(2) is a completely opposite approach to \texttt{CCWS} which throttles warps with low potential of data locality.
Lastly, we show that the synergistic integration of (1) and (2) provides 54\% higher performance than \texttt{CCWS}.

\section{Acknowledgement}
\label{sec:ack}
This research is mainly supported by NRF 2016R1C1B2015312. This work is also supported in part by IITP-2017-2017-0-01015, NRF-2015M3C4A7065645,
DOE DE-AC02-05CH 11231, and MemRay grant (2015-11- 1731). Nam Sung Kim is supported in part by NSF 1640196 and SRC/NRC NERC 2016-NE-2697-A. Jie Zhang and Shuwen Gao contributed equally to this work. Myoungsoo Jung is the corresponding author.

\bibliographystyle{IEEEtran}
\bibliography{ref}

\end{document}